\documentclass[showpacs,aps,twocolumn]{revtex4}
\usepackage[utf8]{inputenc}
\usepackage{epsfig}
\usepackage{graphicx}
\usepackage{amsmath,amssymb,amsfonts}
\usepackage{array}
\usepackage{url}
\usepackage{hyperref}
\usepackage{multirow}
\usepackage{float}
\usepackage{textcomp} 
\usepackage{xspace}
\usepackage[usenames,dvipsnames]{color}
\newcommand{\bef}{\begin{figure}}
\newcommand{\eef}{\end{figure}}
\newcommand{\bc}{\begin{center}}
\newcommand{\ec}{\end{center}}

\newcommand{\be}{\begin{equation}}
\newcommand{\ee}{\end{equation}}
\newcommand{\bea}{\begin{eqnarray}}
\newcommand{\eea}{\end{eqnarray}}

\def\ba{\begin{eqnarray}}
\def\ea{\end{eqnarray}}

\begin{document}

\title{Multiplicity, transverse momentum and pseudorapidity dependence of open-heavy flavored hadron production in proton+proton collisions at $\sqrt{s}$= 13 TeV using PYTHIA8}
\author{Bhagyarathi Sahoo}
\author{Suman Deb}
\author{Raghunath Sahoo\footnote{Corresponding author email: Raghunath.Sahoo@cern.ch}}
\affiliation{Department of Physics, Indian Institute of Technology Indore, Simrol, Indore 453552, India}
\begin{abstract}

Recently, with the upgradation of LHC, it is realized that study of heavy-flavored hadrons, namely $\Lambda_{c}^{+}$ and $\rm{D}^{0}$ in hadronic collisions, could reveal the possibility of thermalization of charm quarks. With this motivation, we study the production dynamics of these hadrons using a pQCD-inspired Monte Carlo event generator called PYTHIA8 in proton+proton collisions at $\sqrt{s}$ = 13 TeV.  The production dynamics of these hadrons are analyzed through charged-particle multiplicity, transverse momentum, and pseudorapidity. With the help of the established PYTHIA8 tunes to mimic the behavior of flow-like features, we investigated the variation of effective temperature and degree of non-extensivity using thermodynamically consistent non-extensive Tsallis statistics. We further attempted to establish a connection between the initial state and final state by estimating the correlation between the number of multi-partonic interactions ($\rm{n}_{MPI}$) with the Knudsen number. 
\end{abstract}
\date{\today}

\maketitle

\section{Introduction}
\label{intro}

Heavy-flavor quarks (charm and beauty) are mostly produced in the early stages of ultra-relativistic high-energy hadronic and nuclear collisions by hard parton-parton scattering processes. These processes predominantly include the production of heavy-flavor quarks through gluon fusion ($gg\rightarrow c\bar{c}/b\bar{b}$) and quark-antiquark annihilation ($q\bar{q}\rightarrow c\bar{c}/b\bar{b}$) at the leading order (LO) in the Quantum Chromodynamics (QCD). Although the next-to-leading order (NLO) perturbative processes ($q\bar{q}\rightarrow c\bar{c}g/b\bar{b}g$, $qg \rightarrow c\bar{c}q/b\bar{b}q$, and $gg\rightarrow c\bar{c}g/b\bar{b}g$) have significant contribution to the production of heavy quarks at the LHC energies~\cite{Norrbin:2000zc}. The hadrons formed from these quarks, both hidden and open, charmed and beauty flavors are considered a vital probe to understand the dynamics of systems formed in both the hadronic and nucleus-nucleus collisions. Starting from their suppression in heavy-ion collisions (HIC) to comprehending the production mechanism in proton+proton (pp) collisions through understanding cold nuclear effect (CNM) in proton on lead collisions, heavy flavored hadrons play a major role. Moreover, because of their heavy mass and larger relaxation time compared to the lifetime of quark-gluon plasma (QGP), a deconfined equilibrated state of quarks and gluons, they experience the entire phase-space evolution process and consequently carry information about the medium. In recent times, both experimental results like strangeness enhancement~\cite{Adam:2017}, ridge-structure~\cite{Li:2011, Khachatryan:2011}, etc., and theoretical results~\cite{Deb:2020, Aditya:2021} of pp collisions have brought a wave of excitement to the particle physics community, which are conventionally considered as heavy-ion-like QGP signatures. These results compel us to reinvestigate the traditional understanding of using pp collisions as a benchmark to study the formation of a QGP medium in HIC. In this context of better understanding the dynamics of pp collisions, heavy-flavored hadrons are believed to be one of the best tools. This is because their dynamics differ from the light-flavored hadrons, as heavy quarks are formed mostly by the perturbative hard-Quantum Chromodynamic (pQCD) processes in contrast to the light quarks, which are produced by non-perturbative Quantum Chromodynamic processes.\\

The production cross-section of heavy flavour hadrons is calculated by several perturbative calculations such as, the General-Mass Variable-Flavour-Number Scheme (GM-VFNS~\cite{Kniehl:2005mk, Kniehl:2012ti}) and Fixed-Order Next-to-Leading-Log (FONLL~\cite{Cacciari:1998it, Cacciari:2012ny})  at next-to-leading order with next-to-leading-log resummation in a wide range of $p_{T}$ range. However, no calculations are available in the latter approach for baryons due to the lack of knowledge of the fragmentation function of charm quarks into baryons. Furthermore, the D meson production cross-section calculations are also available in the $k_{T}$ factorization approach~\cite{Maciula:2013wg}, transverse-momentum dependent factorization approach (TMD)~\cite{Boer:2012bt, Ma:2012hh}, and in the color-glass condensate model (CGC)~\cite{Ma:2014mri}. The production cross-section of heavy flavor hadrons is measured in hadronic and nuclear collisions for various center of mass energies by the ALICE detectors~\cite{ALICE:2020wla, Acharya,ALICE:2023sgl, ALICE:2023wbx, ALICE:2019nxm, ALICE:2020wfu, ALICE:2018hbc, ALICE:2021npz}. Recently, the multiplicity dependence as well as $p_{T}$ differential analysis is performed in ALICE for heavy flavor hadrons in pp collisions at $\sqrt{s} = $ 13 TeV~\cite{ALICE:2021npz}.

In addition, various phenomenological models~\cite{Deb:2018qsl, Ferreiro:2012fb, Kopeliovich:2013yfa} attempt to explain the behavior of relative J$/\psi$ yield as a function of relative charged-particle multiplicity measured in pp collisions at $\sqrt{s} = $ 13 TeV in ALICE experiment at the LHC~\cite{ALICE:2012pet, ALICE:2017wet}. One of these phenomenological models is PYTHIA8, a pQCD-inspired model that could be used to study the dynamics of heavy-flavored production in small systems concerning the LHC energies. The hadronization in the Monte-Carlo (MC) generators PYTHIA8 is based on the formation of strings. It is seen that PYTHIA8, with the help of its tunes, namely multi-parton interactions (MPIs) and color reconnection (CR), could mimic the behavior of flow-like features such as bump structure in baryon-to-meson ratio with transverse momentum ($p_{\rm{T}}$), mass dependant evolution of mean $p_{\rm{T}}$ with scaled event multiplicity~\cite{OrtizVelasquez:2013ofg}. These features of PYTHIA8 have been exploited in literature to study the thermodynamical quantities in pp collisions \cite{Deb:2019yjo,Deb:2020ezw}. To have a comprehensive prospect on the dynamics of heavy flavored baryon and meson, we have studied the $p_{\rm{T}}$-spectra of $\Lambda_{c}^{+}$ and $\rm{D}^{0}$ using PYTHIA8. The choice of these two particles is done keeping in mind the baryon-to-meson ratios in the charm sector, which is of importance in Run 3 of the LHC. It is worth recalling here that the possibility of thermalization in small systems was studied in 1953 by Landau in the context of collisions of small systems~\cite{Landau:1953, Landau:1955, Landau:1956, Landau:1965}. This shows the importance of the subject matter and its implications for LHC energies.\\

The statistical approach gives a better description of dealing with a complex system formed in the hadronic and nuclear collisions. The study of final state transverse momentum spectra dispenses knowledge about the kinetic freeze-out processes in such collisions. With this motivation, we have analyzed the $p_{\rm{T}}$-spectra of $\Lambda_{c}^{+}$ and $\rm{D}^{0}$ to extract thermodynamic information in a wide $p_{\rm{T}}$-range using a thermodynamically consistent non-extensive distribution function, called the Tsallis distribution \cite{Cleymans:2011in, Cleymans:2012ya, Wong:2013sca, Khuntia:2018znt, Tsallis:1987eu, Bhattacharyya:2015hya}. This distribution function explains the transverse momentum spectra of final state particles in both hadronic and HIC collisions at a larger $p_{\rm{T}}$-range very well while taking care of both the aspects of particle production, namely, the pQCD-based and non-pQCD-based processes. Further, it contains two important parameters: effective temperature (T) or Tsallis temperature and non-extensive parameter (q). This non-extensive parameter indicates the degree of deviation of the system from an equilibrium state, which is correlated with the Tsallis temperature. The effective temperature at the freeze-out carries the information of the random thermal motion in terms of the thermal temperature and the collective effects constituting a common radial flow velocity~\cite{Ghosh:2014eqa, Khuntia:2018znt}. Since charged-particle multiplicity is a good replacement for centrality in the pp system and acts as a proxy for the number of constituents formed in a system after the collision, we have studied the production dynamics of heavy-flavor particles with charged-particle multiplicity. Through various studies~\cite{ Rath:2019cpe, Patra:2020gzw}, it is understood that different regions of $p_{\rm{T}}$-spectra could shed more light on the origin of particle hadronization while traversing through the medium. With this in mind, we have also investigated the variation of particle production dynamics using transverse momentum spectra. Further, in the HIC, since the midrapidity regions are mostly dominated by gluons while forward rapidity by the constituent quarks~\cite{Wang:2006xq, Wang:2007zv}, it is interesting to study the rapidity dependence of particle production, which is explored here.\\

One of the purposes of the present work is to study the possibility of a thermalization-like effect on heavy-flavored hadrons, viz., $\Lambda_{c}^{+}$ and $\rm{D^{0}}$ in pp collisions by calculating some of the thermodynamical quantities like the squared speed of sound ($ c_{s}^{2}$), isothermal compressibility ($\kappa_{T} $) and the Knudsen number ($K_{n}$). The squared speed of the sound describes the interaction strength of the particles in a medium. While isothermal compressibility describes the compression nature of the medium, the Knudsen number describes the hydrodynamical behavior of the system. The fluctuation in Tsallis temperature is related to the non-extensive parameter (q) \cite{Wilk:2012zn} and both these parameters are particle species dependent \cite{Khuntia:2018znt}. In this regard, we have analyzed the statistical correlation between T and q to investigate their dependence on the mass of the particle. Since we are using PYTHIA8 for the present work, we have access to the mean number of multi-partonic interactions ($\langle n_{\rm{MPI}}\rangle$), which gives information on the interaction at the partonic level. The calculation of the inverse of the Knudsen number at the hadronic-phase boundary provides information on the number of hadronic interactions at the phase boundary. It would be interesting to understand the correlation dynamics between these two numbers, as they connect the initial state with the final state of the system.  Heavy-flavored baryons and mesons could be a better probe as they witness the whole system's space-time evolution.\\

This paper is organized as follows. After the introduction and definition of the physics problem in hand, in the present section~\ref{intro}, we move forward in giving a brief detail of event generation using PYTHIA8 and the analysis methodology in section~\ref{eventgen}. Section~\ref{result} includes the results obtained from the analysis of $p_{\rm{T}}$-spectra  for $\Lambda_{c}^{+}$ and $\rm{D^{0}}$ in pp collisions. Finally, we conclude with a summary of the important findings in section~\ref{summary}.

\section{EVENT GENERATION AND ANALYSIS METHODOLOGY}
\label{eventgen} 
\subsection{Event Generation}

Event Generators are the most effective substitutes for realistic experiments involving hadronic and nucleus-nucleus collisions (AA). These generators are based on the Monte Carlo techniques for simulating the collisions using physics processes to better understand the nature of the events. Based on the kind of colliding species involved and underlying known physics processes, there are several event generators like AMPT~\cite{Lin:2004en} (for nucleus-nucleus collisions), PYTHIA (for both pp and AA collisions), EPOS LHC~\cite{Pierog:2013ria} (for hadronic and nuclear collisions), etc. Among them, PYTHIA is one of the widely used event generators for simulating ultra-relativistic collisions among particles like electron-electron, electron-positron, proton-proton, and proton-antiproton. It is highly successful in explaining many experimental results at the LHC. PYTHIA involves many physics mechanisms like hard and soft interactions, parton distributions, initial and final state parton showers, multipartonic interactions, string fragmentation, color reconnection and resonance decays, rescattering, and beam remnants~\cite{Sjostrand}.\\

In this study, we have used PYTHIA 8.235 to generate pp collisions at $\sqrt{s}$ = 13 TeV with 4C Tune (Tune:pp = 5)~\cite{Corke}.  Subsequent Multi-Partonic Interaction (MPI) process is one of the main advantages of PYTHIA 8.235 over PYTHIA6 along with impact parameter dependence of collision, which allows heavy-flavor quarks production through 2 $\rightarrow$ 2 hard subprocesses. A detailed explanation of all physics processes involved in PYTHIA 8.235 can be found in Ref.~\cite{manual}.\\

For our study, we contemplate inelastic, non-diffractive simulated events. So in the total scattering cross-section, only the non-diffractive component of all hard QCD processes (HardQCD:all=on) will contribute. Hard processes involve the production of heavy quark flavors through all hard-QCD $2 \rightarrow 2$ scattering processes. It involves pair creation [q$\bar{q}$ (gg) $\rightarrow$ c$\bar{c}$], flavour excitation, and gluon splitting processes~\cite{manual}. We have considered mode two of color reconnection (ColourReconnection:mode = 2) along with MPI (PartonLevel:MPI = on). The choice of CR mode 2 of PYTHIA8 is because it qualitatively describes the experimental data of $\Lambda_{c}^{+}/D^{0}$ ratios, however, it underestimate the $p_{T}$-spectra of prompt $\Lambda_{c}^{+}$ at low $p_{T}$~\cite{ALICE:2020wla}. The data are simulated considering color reconnection on (ColourReconnection:reconnect = on) and color reconnection off (ColourReconnection:reconnect = off). To avoid the divergences of QCD processes in the limit $p_{\rm{T}} \rightarrow 0 $  a transverse momentum cut of $ p_{\rm{T}} \geq 0.5 $ GeV/c  (PhaseSpace:pTHatMinDiverge = 0.5) is taken along with a $p_{\rm T}$ minimum phase space cut-off of 2.0 (PhaseSpace:pTHatMin = 2.0). For the production of $\Lambda_{c}^{+}$ and $\rm{D^{0}}$, we use Charmonium:all flag (Charmonium:all = on) in the simulation~\cite{Shao, Caswell, Bodwin} through NRQCD framework. Study of  $\Lambda_{c}^{+}$ and $\rm{D^{0}}$ are done at mid and forward rapidities. $\Lambda_{c}^{+}$ ($\rm{D^{0}}$)  decays via $p + K^{-} +\pi^{+} $~\cite{Acharya} ($K^{-} + \pi^{+}$~\cite{Hamon}) because of high branching ratio of these modes. This analysis is performed by generating 350 million events in pp collisions at  $\sqrt{s}$ = 13 TeV for all the rapidities discussed here.\\

To check the compatibility of PYTHIA8 with experimental data, we have used the same tuning as used in one of our previous works described in Ref~\cite{Deb:2020ige}, where we have compared the experimental data with PYTHIA8 simulated data for $\rm{D^{0}}$.\\


\subsection{ Analysis Methodology}
\label{formalism}
   \begin{figure*}[!htb]
  \bc

       \includegraphics[scale=0.33]{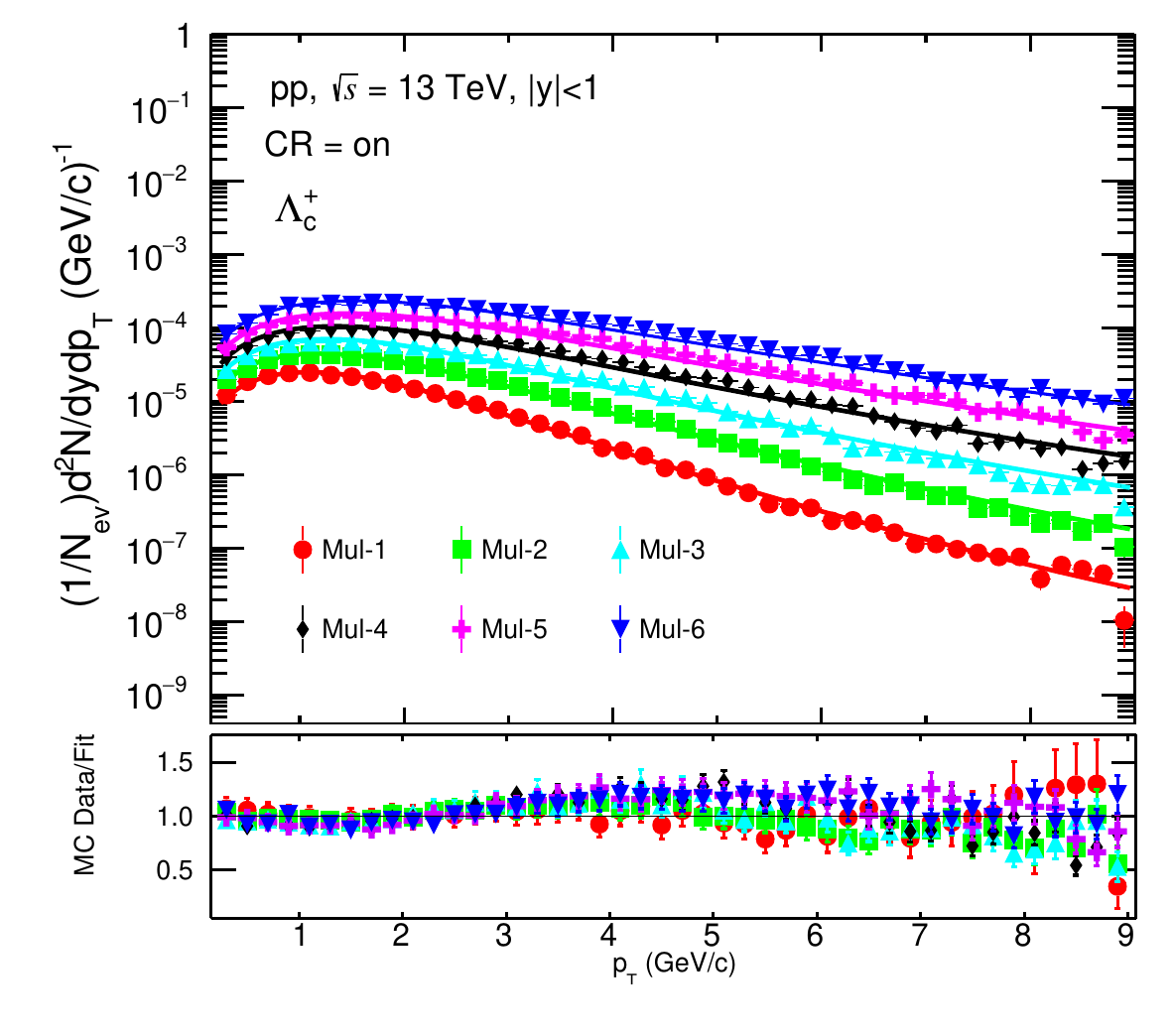}
        \includegraphics[scale=0.33]{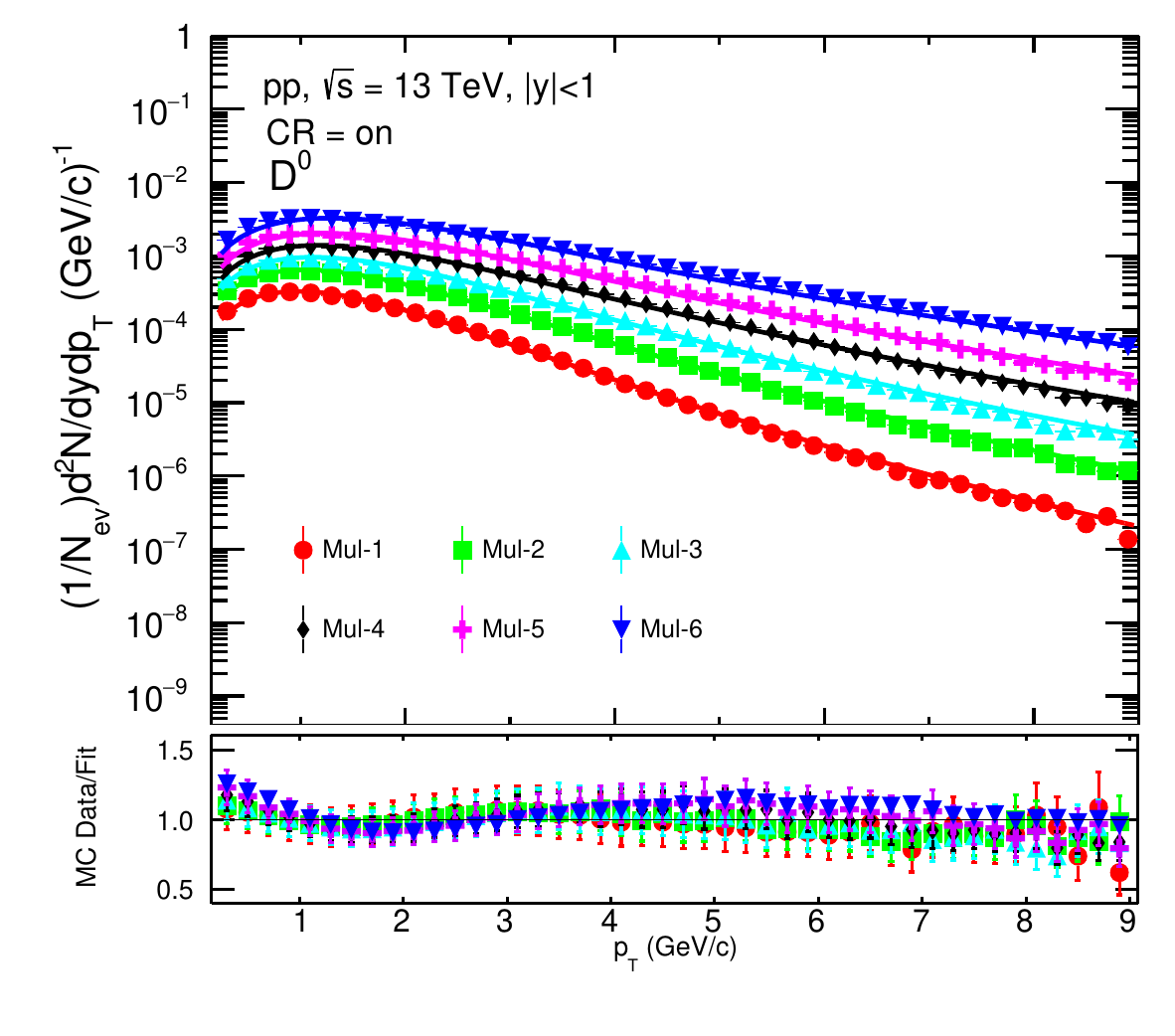}
       \includegraphics[scale=0.33]{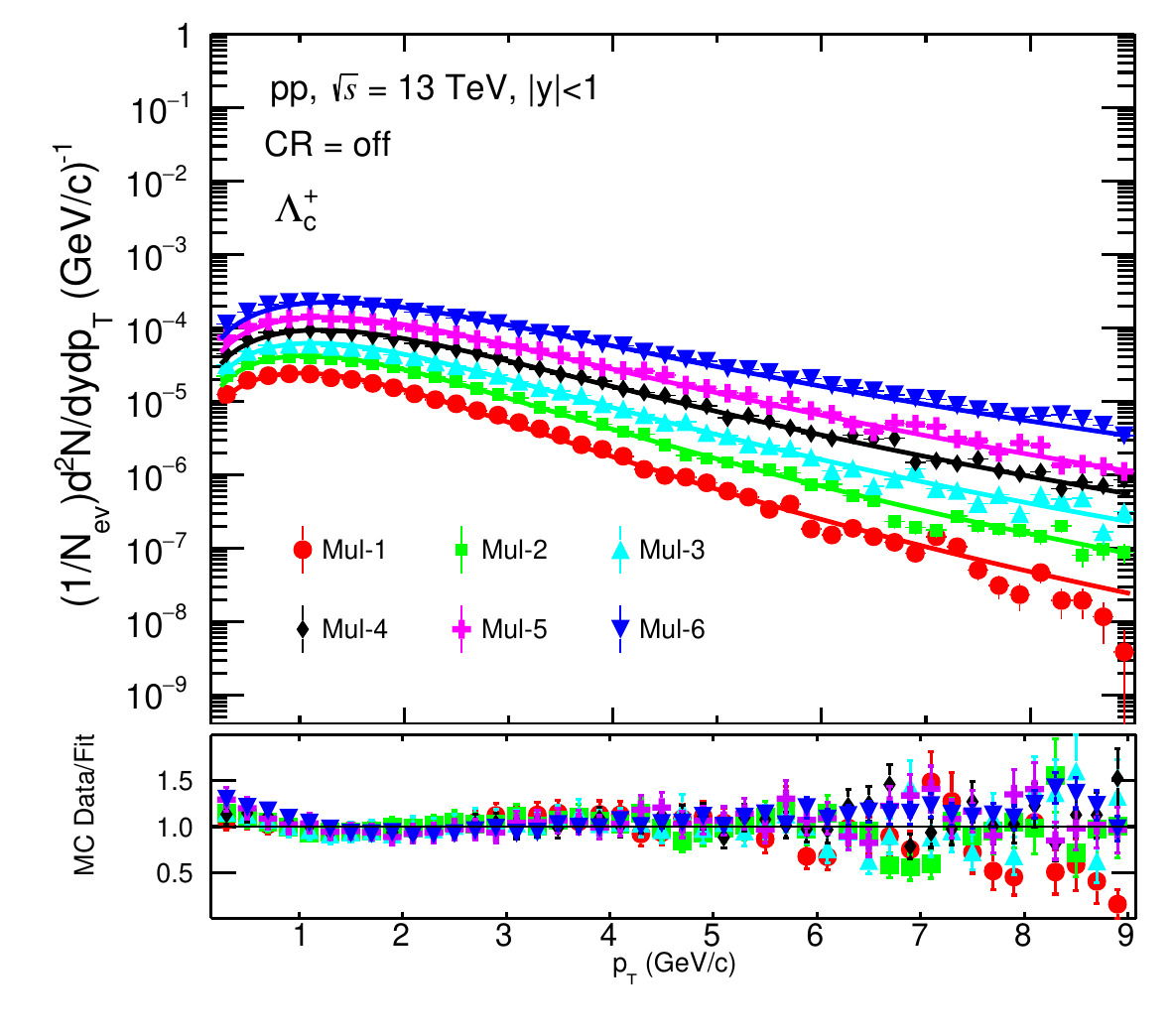}
      \includegraphics[scale=0.33]{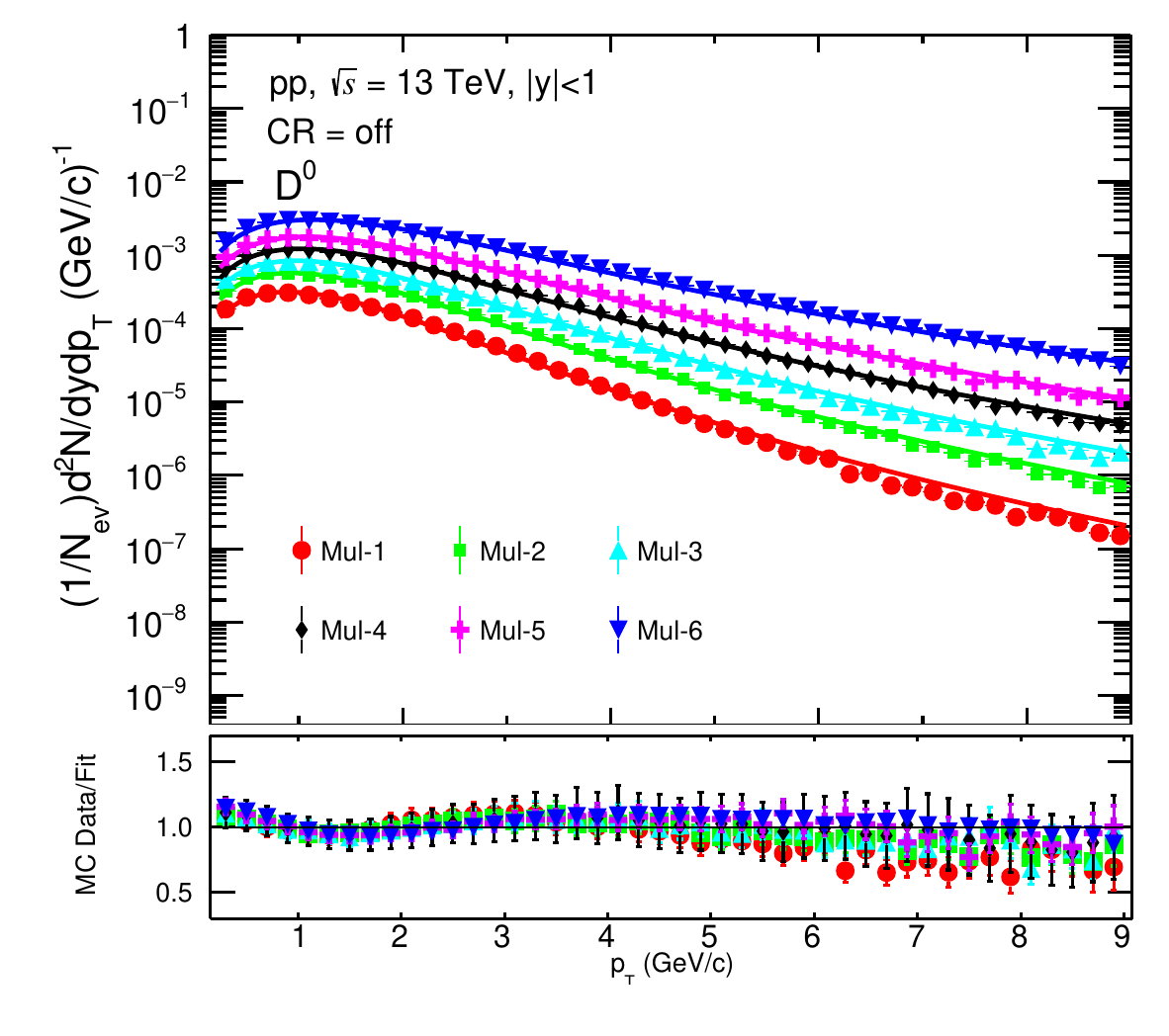}

   \caption{(Color online) {Upper (lower) panel shows the fitting of simulated $p_{\rm{T}}$-spectra of $\Lambda_{c}^{+} $} and $\rm{D^{0}}$ from PYTHIA8 using Tsallis distribution function for CR = on (CR = off) in different multiplicity classes as shown in Table ~\ref{table:mult_info}.}
 
  \label{fig:1}
\ec
\end{figure*}

The $p_{\rm{T}}$-spectra of produced particles by a system in thermal equilibrium follow a Boltzmann type of distribution, given as~\cite{STAR:2006nmo}
\begin{equation}
E\frac{d^{3}N}{dp^{3}} = C \;\exp{ \left( \frac{-\sqrt{m^2 + p_{\rm T}^{2}}}  {T} \right) },
\end{equation}

where C is the normalization constant and $T$ is the temperature. It is observed that at LHC/RHIC, the low-$p_{\rm{T}}$ part ($< 3 $ GeV/c) is well explained by the Boltzmann-Gibbs Blast Wave model taking radial flow into account in pp, pA, and AA collisions~\cite{Ghosh:2014eqa,ALICE:2019hno,ALICE:2013mez}, but the high-$p_{\rm{T}} $ spectra do not follow Boltzmann type distribution~\cite{Bhattacharyya:2015hya}. This could be because of possible pQCD processes, which have a power-law contribution to the spectra. 

To explain the whole range of $p_{\rm{T}}$-spectra, one has to take care of the power-law contribution at high-$p_{\rm{T}}$~\cite{Michael:1977, Michael:1979, Arnison}. Hagedorn proposed an empirical formula to describe the whole $p_{\rm{T}}$ range, in which high-$p_{\rm{T}}$ region follows a power-law function, while the low-$p_{\rm{T}}$ follows an exponential function~\cite{Hagedorn}. A similar kind of observation also can be found with the thermodynamically consistent form of the Tsallis distribution function having a functional form, given by:
\begin{align}
    f(E, \mu, T, q) \equiv \exp_{q}\left(-\frac{E-\mu}{T}\right) \equiv \left( 1+(q-1)\frac{E-\mu}{T}\right)^{-\frac{1}{q-1}} 
    \label{eq_2}
\end{align}

where $ E= \sqrt{p^{2}+m^{2}}$ is the energy of the particle, and $\mu$ is the chemical potential of the system. Here, T is the effective temperature. This effective temperature depends on the particle mass and the collective flow velocity. The intercept in $T = T_{\rm kin} + \frac{1}{2} \rm m \beta_{\rm T}^{2}$ is an alternative method to find out the kinetic freeze-out temperature ($T_{ \rm kin}$) in the low-$p_{\rm T}$ limit. Where m is the rest mass of the particle and $\beta_{\rm T}$ is the average transverse radial flow velocity~\cite{Patra:2020gzw} \\

In Eq.~\ref{eq_2}, q is the non-extensive parameter that measures the degree of
deviation of the system from the thermodynamic equilibrium state. In general q $\geq$ 1. It is also called as entropy index. The entropy index q characterizes the non-extensive properties of the system such as it measures the degree of non-additive nature of the entropy, details of which can be found in Ref.~\cite{Tsallis}. \\

The $\exp_{q}(x)$ has the following form:

\begin{equation}
   \exp_{q}(x) \equiv 
    \begin{cases}
    [1+(q-1)x]^\frac{1}{q-1}  \hspace{0.8cm} \text{if}  \hspace{0.1cm} $$ x  >  0 $$\\
    [1+(1-q)x]^\frac{1}{1-q} \hspace{0.8cm} \text{if}  \hspace{0.1cm} x \leq 0
    \end{cases}
    \label{eq_3}
\end{equation}

where $ x = \frac {E- \mu}{T}$. In the limit, q$\rightarrow$1, Eq.~\ref{eq_3} reduces to an exponential function.

\begin{equation}
 \lim_{q\to1}\exp_{q}(x) \rightarrow \exp(x)
\end{equation}

Therefore, under the limit q$\rightarrow$1, the non-extensive Tsallis distribution function defined in Eq.~\ref{eq_2} reduces to the extensive Boltzmann-Gibbs (BG) distribution function
\begin{equation}
        f_{q \rightarrow 1}(E, \mu, T, q) \equiv   f_{BG} (E, \mu, T) \equiv \exp\left(-\frac{E-\mu}{T}\right)
 \label{fatq1}
\end{equation}

Consequently, the invariant yield in terms of Eq.~\ref{eq_2}  could be expressed as

\begin{equation}
 E\frac{d^{3}N}{dp^{3}} = gVE \frac{1}{(2\pi)^{3}}\left(1+(q-1)\frac{(E-\mu)}{T}\right)^{\frac{-q}{q-1}}
 \label{eq_5}
 \end{equation}
 where g is the degeneracy factor, V is the volume of the system which depends on the q.\\
 
 In terms of rapidity (y), transverse mass ($m_{T}$), and transverse momentum ($p_{\rm{T}}$), Eq.~\ref{eq_5} becomes

 \begin{equation} 
 \frac{d^{2}N}{dp_{\rm{T}}dy} =  gV\frac{p_{\rm{T}}m_{T} \cosh y}{(2\pi)^{2}}\left(1+(q-1)\frac{(m_{T}\cosh y-\mu)}{T}\right)^{\frac{-q}{q-1}} 
 \label{eq_6}
\end{equation} 

where, $m_{T} = \sqrt{p_{\rm{T}}^{2}+m^{2}}$ is the transverse mass. \\

At LHC energies, the chemical potential is assumed to be nearly zero (${\mu}\simeq{0}) $ and at midrapidity (y=0), Eq.~\ref{eq_6} reduces to :
\begin{equation}
\frac{d^{2}N}{dp_{\rm{T}}dy} = gV\frac{p_{\rm{T}}m_{T}}{(2\pi)^{2}}\left(1+(q-1)\frac{m_{T}}{T}\right)^{\frac{-q}{q-1}}
\end{equation}

The thermodynamic quantities like number density (n), energy density ($\epsilon$), and pressure (P) can also be calculated in terms of the Tsallis distribution function defined in Eq.~\ref{eq_2} to search for the thermal behavior of the discussed particles in each multiplicity class. These quantities as expressed as follows,

\begin{equation}
n = g\int \frac{d^{3}p}{(2\pi)^{3}}f^{q}
\label{number density}
 \end{equation}
 
 \begin{equation}
             \epsilon = g \int \frac{d^{3}p}{(2\pi)^{3}}Ef^{q} 
             \label{energy density}
 \end{equation}
\begin{equation}
    P = g\int \frac{d^{3}p}{(2\pi)^{3}}\frac{p^{2}}{3E}f^{q}
    \label{Pressure}
\end{equation}  
where, 
\begin{equation}
f^q \equiv f^q(E, \mu, T, q) \equiv \left( 1+(q-1)\frac{E-\mu}{T}\right)^{-\frac{q}{q-1}} 
\label{fq}
\end{equation}
By using the above thermodynamic quantities, one can obtain the crucial indicators of thermalization like the squared speed of sound ($c_{s}^{2}$), isothermal compressibility ($\kappa_{T}$), and Knudsen number ($ K_{n}$), which are of interest for the present work.\\

The squared speed of sound is defined as the change in pressure of the system with respect to a change in the energy density at constant entropy density and number density. Mathematically,

\begin{equation}
     c_{s}^{2} =\left(\frac{ \partial{P}}{\partial\epsilon}\right)_{s,n}
     \label{eq_11}
\end{equation}

It can also be defined as,
\begin{equation}
     c_{s}^{2} =\frac{\left(\frac{ \partial{P}}{\partial{T}}\right)}{\left(\frac{\partial\epsilon}{\partial{T}}\right)}
\end{equation}

It describes the hydrodynamical evolution and equation of the state of the thermal medium. It helps us to characterize the interaction strength of the QCD medium formed after the collision, i.e., whether it is strongly interacting or not or how much it differs from the ideal massless non-interacting gas. It is inversely proportional to the compressibility of the fluid.\\

The isothermal compressibility ($ \kappa_{T}$) is defined as the change in volume of the system with a change in pressure at a constant temperature.
\begin{equation}
    \kappa_{T} = -\frac{1}{V}\left(\frac{\partial{V}}{\partial{P}}\right)_{T}
    \label{eq_12}
\end{equation}
It is a response function that defines the compression nature and equation of the state of the medium. The value of isothermal compressibility is very sensitive to charged-particle multiplicity fluctuation, temperature, and volume of the system.\\

In terms of multiplicity fluctuations and average number, isothermal compressibility can be defined as~\cite{Sahu:2020swd}.
\begin{equation}
   \textlangle \left(N - \textlangle{N \textrangle} \right)^2 \textrangle = var(N) = \frac{T \textlangle{N \textrangle}^2   \kappa_{T}}{V}
    \label{eqKT1}
\end{equation}
where N is the particle multiplicity. From the basic thermodynamics relation, we have
\begin{equation}
    \textlangle \left(N - \textlangle{N \textrangle}\right)^2 \textrangle = VT \left( \frac{\partial n}{\partial \mu} \right)
    \label{eqKT2}
\end{equation}
Thus, comparing Eqs.~\ref{eqKT1} and ~\ref{eqKT2}, we obtain

\begin{equation}
      \kappa_{T} =\frac{1}{n^2}\left(\frac{\partial n}{\partial \mu} \right)
      \label{KT}
\end{equation}
where,
\begin{equation}
    \frac{\partial n}{\partial \mu} = \frac{gq}{T}\int \frac{d^{3}p}{(2\pi)^{3}}\bigg[1+(q-1)\frac{E - \mu}{T}\bigg]^\frac{1-2q}{q-1}
    \label{delndelmu}
\end{equation}
The baryochemical potential of the system is almost zero at the LHC energies. So, for our studies, we use $\mu$ = 0 in the calculations.\\

Knudsen number ($K_{n}$) is defined as the ratio of the mean free path ($\lambda$) of a particle to a spatial dimension of the system (R)~\cite{Scaria:2022yrz}.
\begin{equation}
           K_{n} = \frac{\lambda}{R} 
           \label{eq_13}
\end{equation}
 It defines the hydrodynamic behavior of the system. When the Knudsen number, $K_{n} >> 1$, this indicates that the system is far from thermodynamic equilibrium, and a smaller value of the Knudsen number, tending to zero, indicates that the system has higher degrees of thermalization and follows a hydrodynamic behavior.\\
 
  In thermodynamics, the mean free path is defined as the average distance traveled by a particle between successive collisions. The mean free path is inversely proportional to the number density times the total scattering cross-section. Mathematically,

 \begin{equation}
     \lambda = \frac{1}{ n \sigma}
 \end{equation}
where $n$ is the number density of the system and is estimated from Eq.~\ref{number density}. The $\sigma$ is the scattering cross-section of the particles. Furthermore, considering hadrons as hard spheres of equal radius $r_h$ with constant cross-section $\sigma = 4\pi r_h^{2}$. In the present study, we consider the hard-core radius $r_h$ = 0.62 fm for $\Lambda_{c}^{+} $ and $r_h$ = 0.2 fm for $\rm{D^{0}}$~\cite{Sahoo:2023vkw}. In addition, the system radius (R) is extracted from one of the Tsallis parameters i.e. the volume parameter (V) by assuming a spherical symmetry of the system. Hence, the radius parameter R is given by, R $\equiv \left[ \frac{3V}{4\pi} \right]^{1/3}$~\cite{Khuntia:2017ite}. This radius parameter varies with the charged particle multiplicity and specific hadronic species. Thus, to calculate the Knudsen number in the present study we consider the mean free path of particle to the system size of the specific hadrons obtained from the Tsallis fit. However, it is noteworthy to mention that the R is not necessarily related to the system size obtained from the experimental HBT analysis, but it is related to the normalization factor in the statistical distribution function used to describe the particle spectra~\cite{Cleymans:2013rfq}.
 
With a brief introduction to the methodology and all the quantities involved in the present analysis, we now move toward the results and discussion section.
 
\section{RESULTS AND DISCUSSION}
\label{result}

\begin{figure*}[!htb]
\bc
  \includegraphics[scale=0.33]{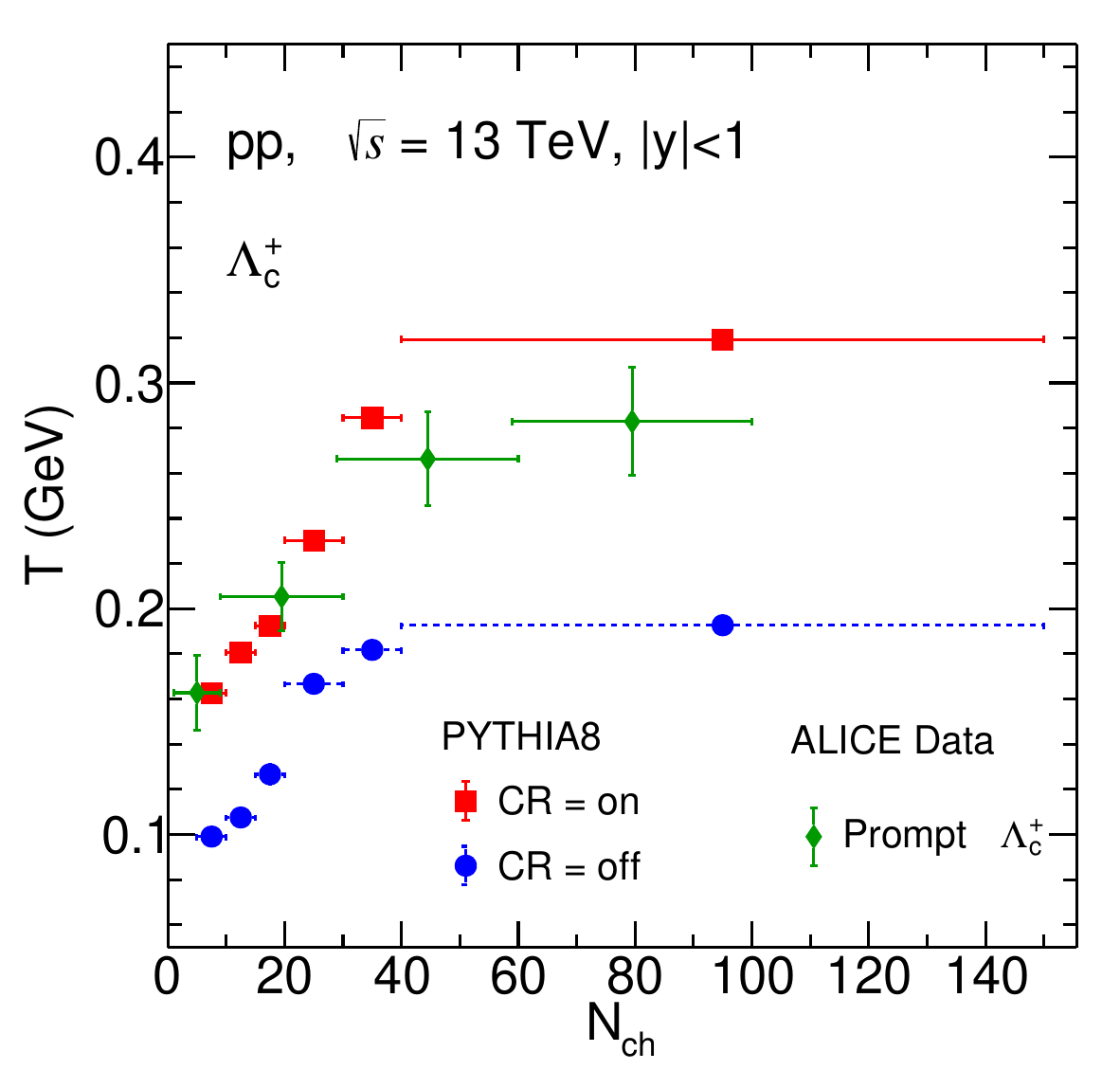}
  \includegraphics[scale=0.33]{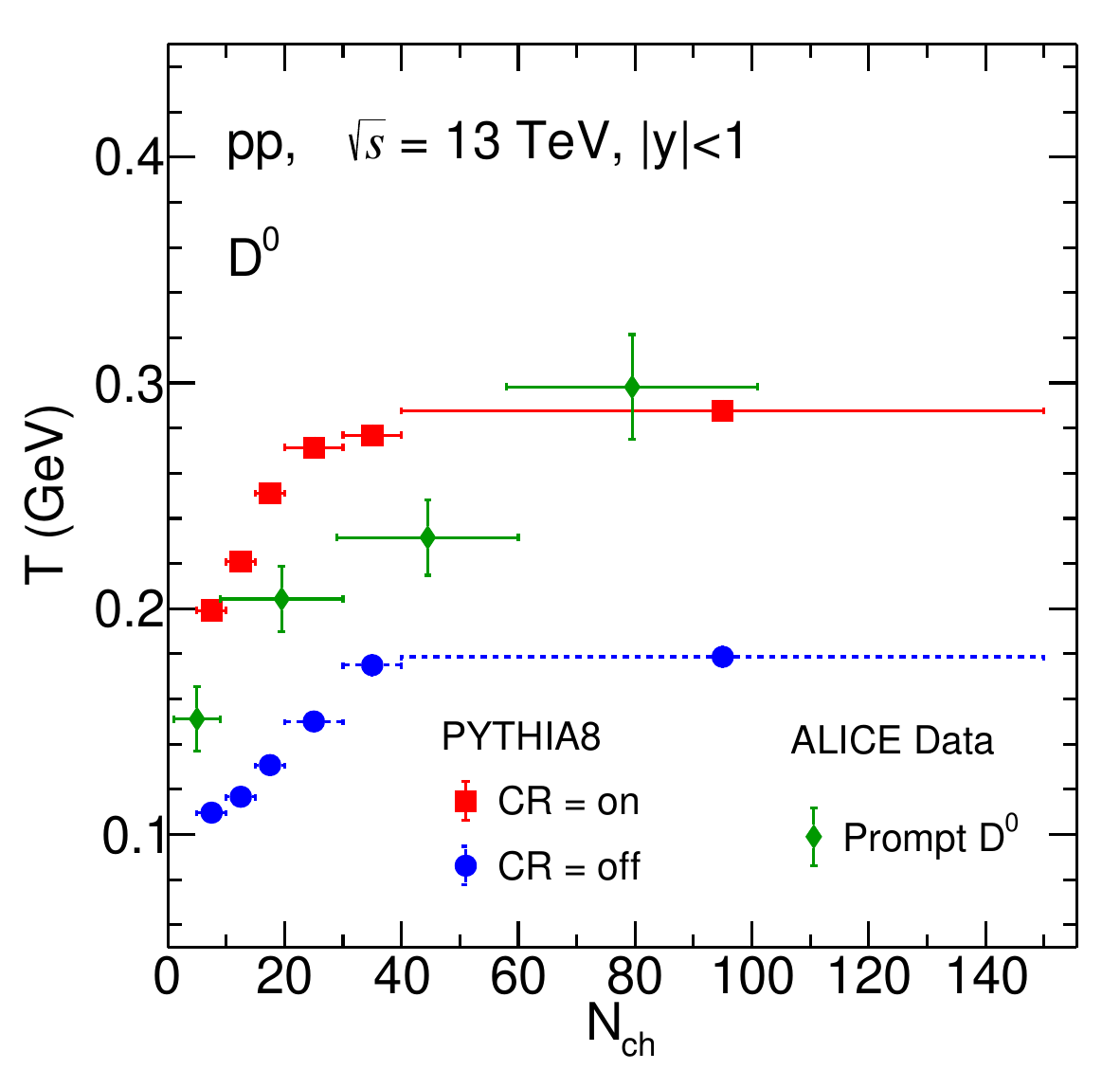}
  \includegraphics[scale=0.33]{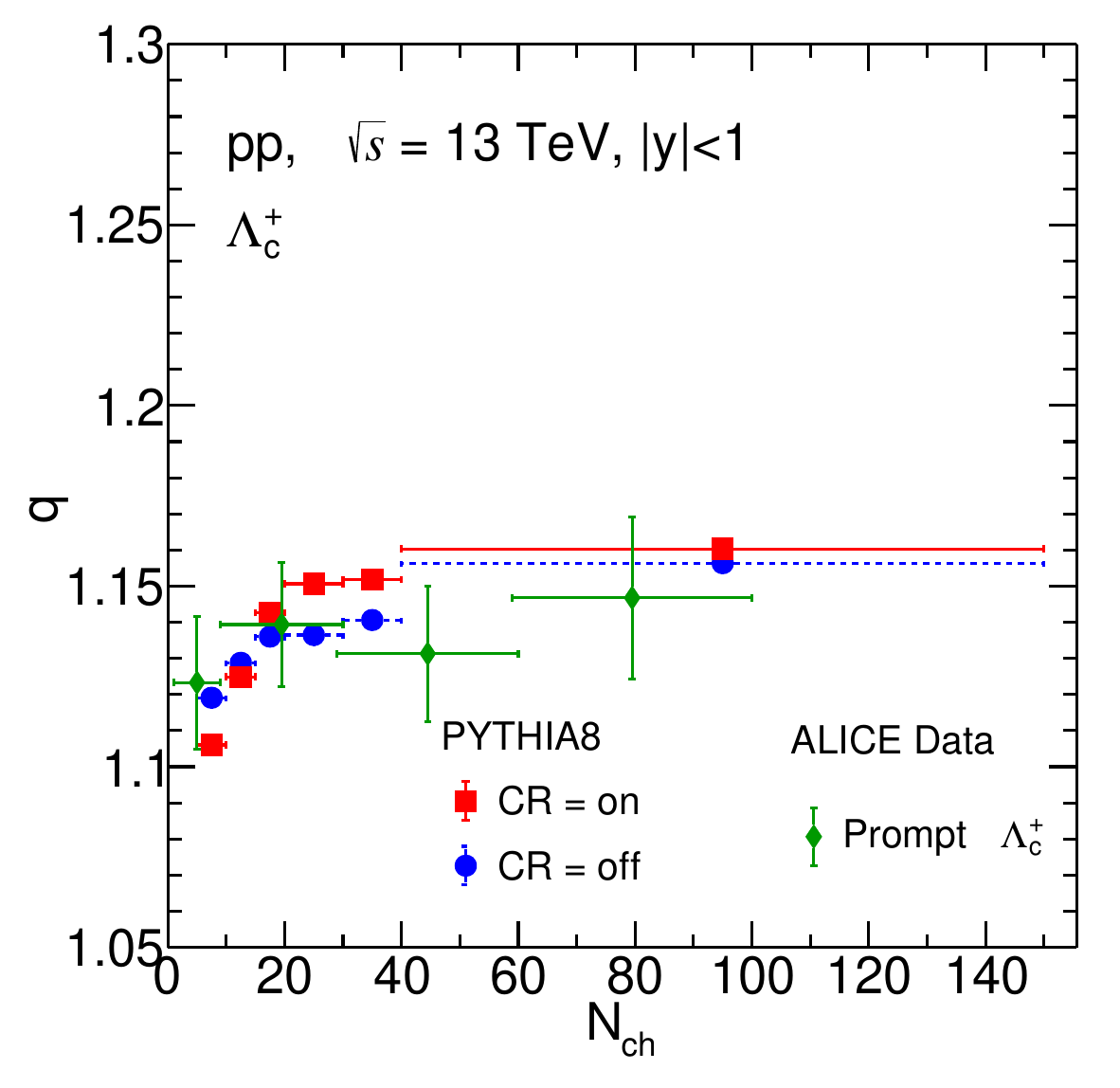}
  \includegraphics[scale=0.33]{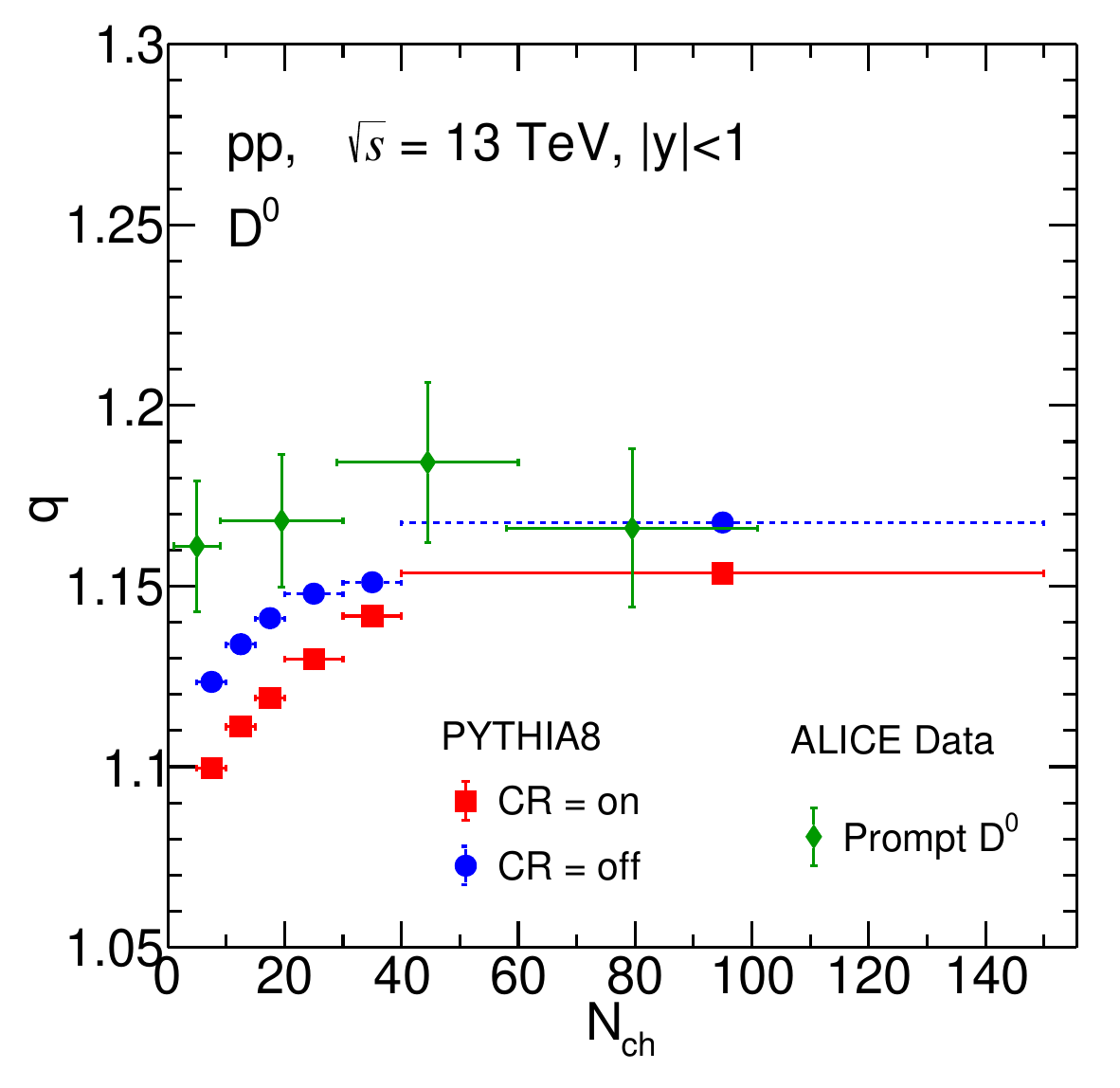}
  
   \caption{(Color online) Upper and lower panels show the Tsallis extracted parameters, T and q with charged-particle multiplicity, respectively by considering both color reconnection modes on and off, for inclusive $\Lambda_{c}^{+} $ and $\rm{D^{0}}$ in pp collisions at $\sqrt{s}$ = 13 TeV using PYTHIA8. The obtained results are compared with the multiplicity dependence of prompt $\Lambda_{c}^{+} $ and $\rm{D^{0}}$ measured by ALICE in pp collisions at $\sqrt{s}$ = 13 TeV, taken from Ref.~\cite{ALICE:2021npz}. The statistical uncertainties associated with the PYTHIA8 results are within the marker size.}
  \label{fig:2}
\ec
\end{figure*}

\subsection{Transverse momentum spectra and Non-extensivity}
\label{pt_spectra}
The $p_{\rm{T}}$-spectra of $\Lambda_{c}^{+}$ and $\rm{D^{0}}$ obtained from the PYTHIA8 simulation of pp collision at $\sqrt{s}$= 13 TeV are fitted with the non-extensive statistical distribution function and are shown in Fig.~\ref{fig:1}.  The deviation of the fitting function from transverse momentum spectra of both $\Lambda_{c}^{+}$ and $\rm{D^{0}}$  are shown in the bottom panels. The fitting parameters, viz., effective temperature (T), and non-extensive parameter (q), were obtained as a function of charged-particle multiplicity. The charged-particle multiplicity classes used in the present analysis are taken from Ref~\cite{Deb:2020ige} and are shown in Table.~\ref{table:mult_info}.

\begin{table*}[htbp]
\caption[p]{Charged-particle multiplicity classes ($N_{ch}$) ($|\eta|$ $<$ 0.8) corresponding to different events}
\label{table:mult_info}
\centering
\begin{tabular}{c|c|c|c|c|c|c|c|c|c|c|c|}
\hline
\multicolumn{2}{|c|}{${\bf Multiplicity Class}$}&Mul-1&Mul-2&Mul-3&Mul-4&Mul-5&Mul-6\\
\hline

\multicolumn{2}{|c|}{ $  \bf {N_{ch}} $} &5-10&10-15&15-20&20-30&30-40&40-150\\
\hline
\end{tabular}
 \end{table*}
 
\begin{figure*}[!htb]
\bc
\includegraphics[scale=0.33]{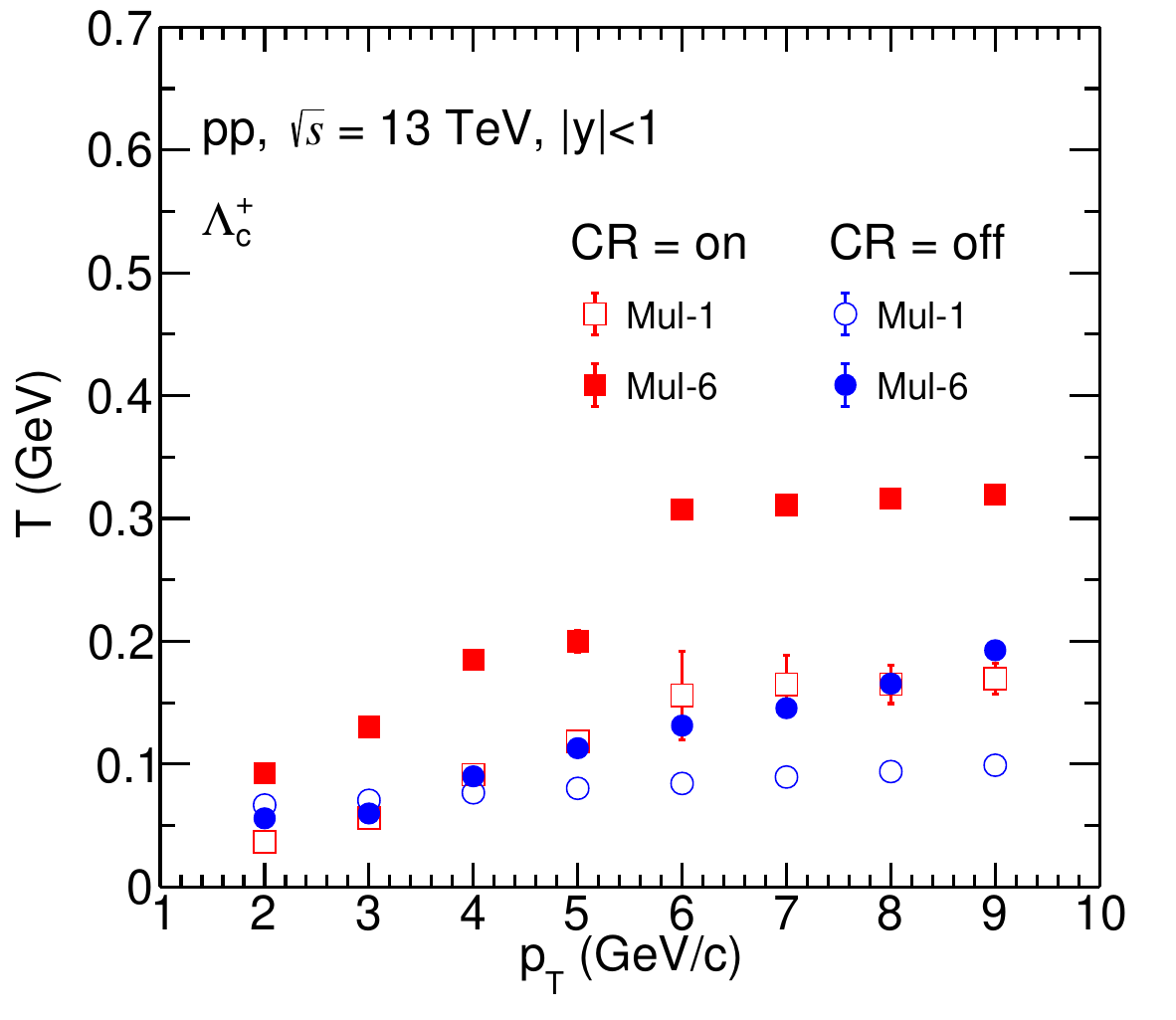}
  \includegraphics[scale=0.33]{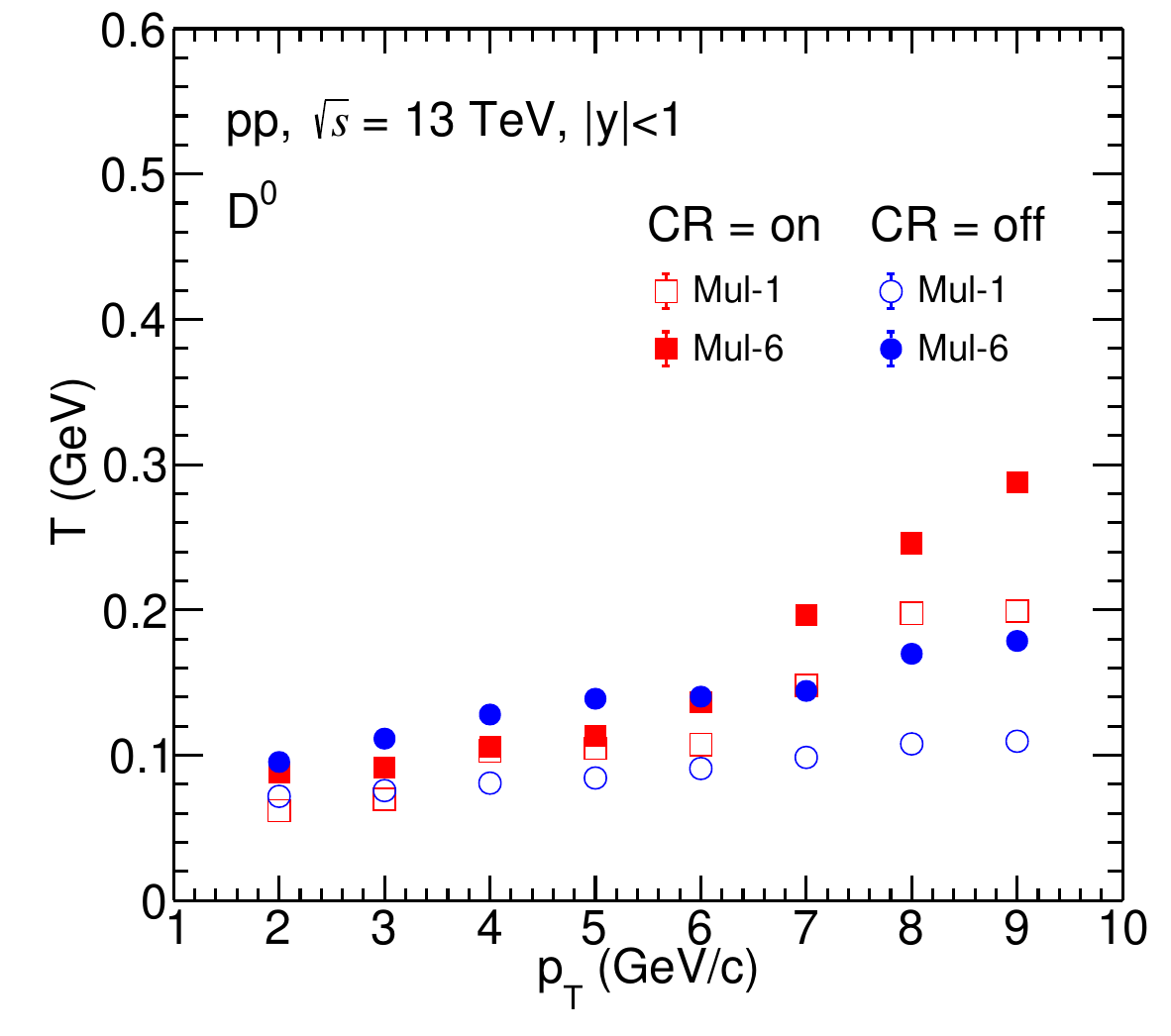} 
  \includegraphics[scale=0.33]{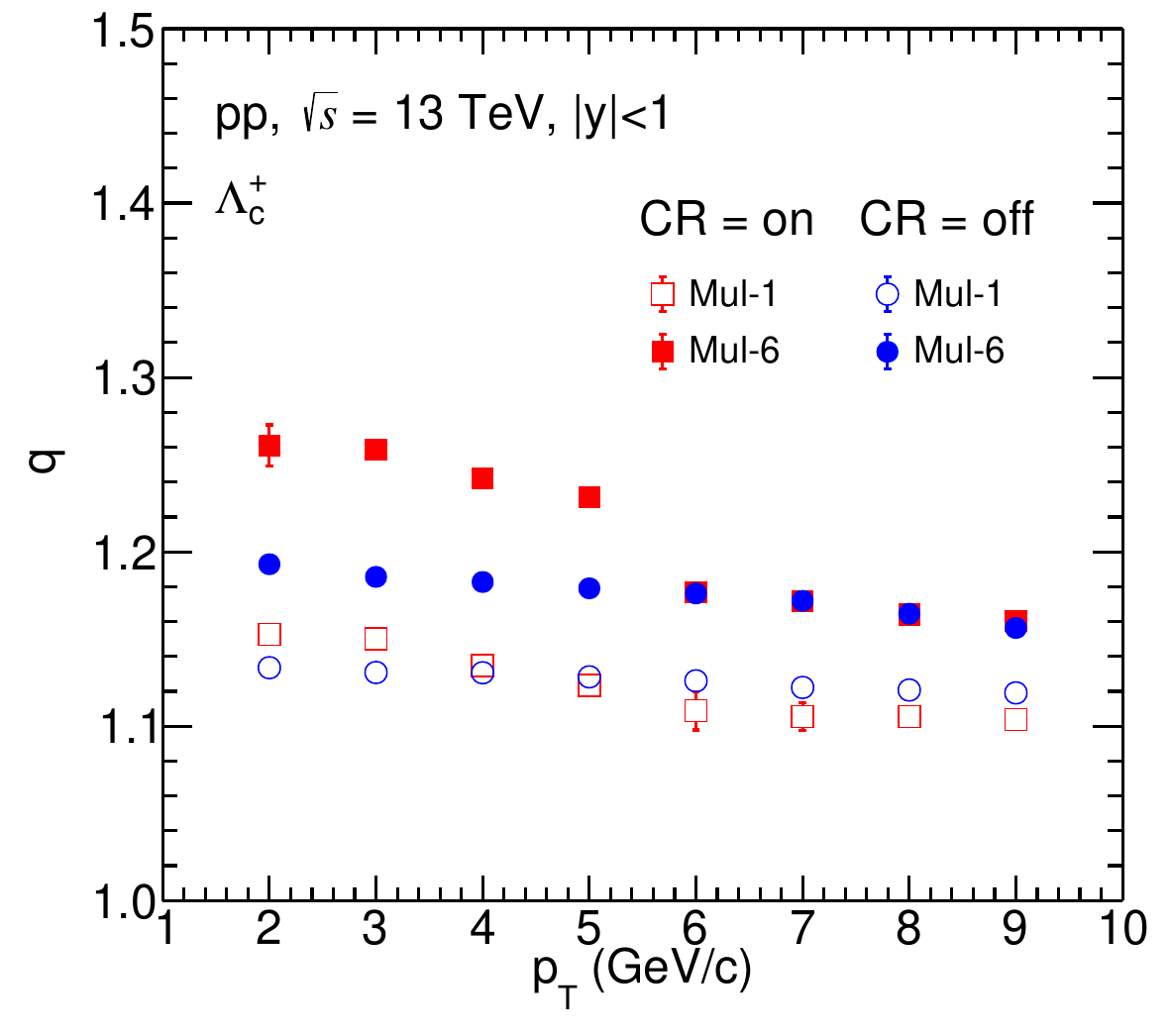}
 \includegraphics[scale=0.33]{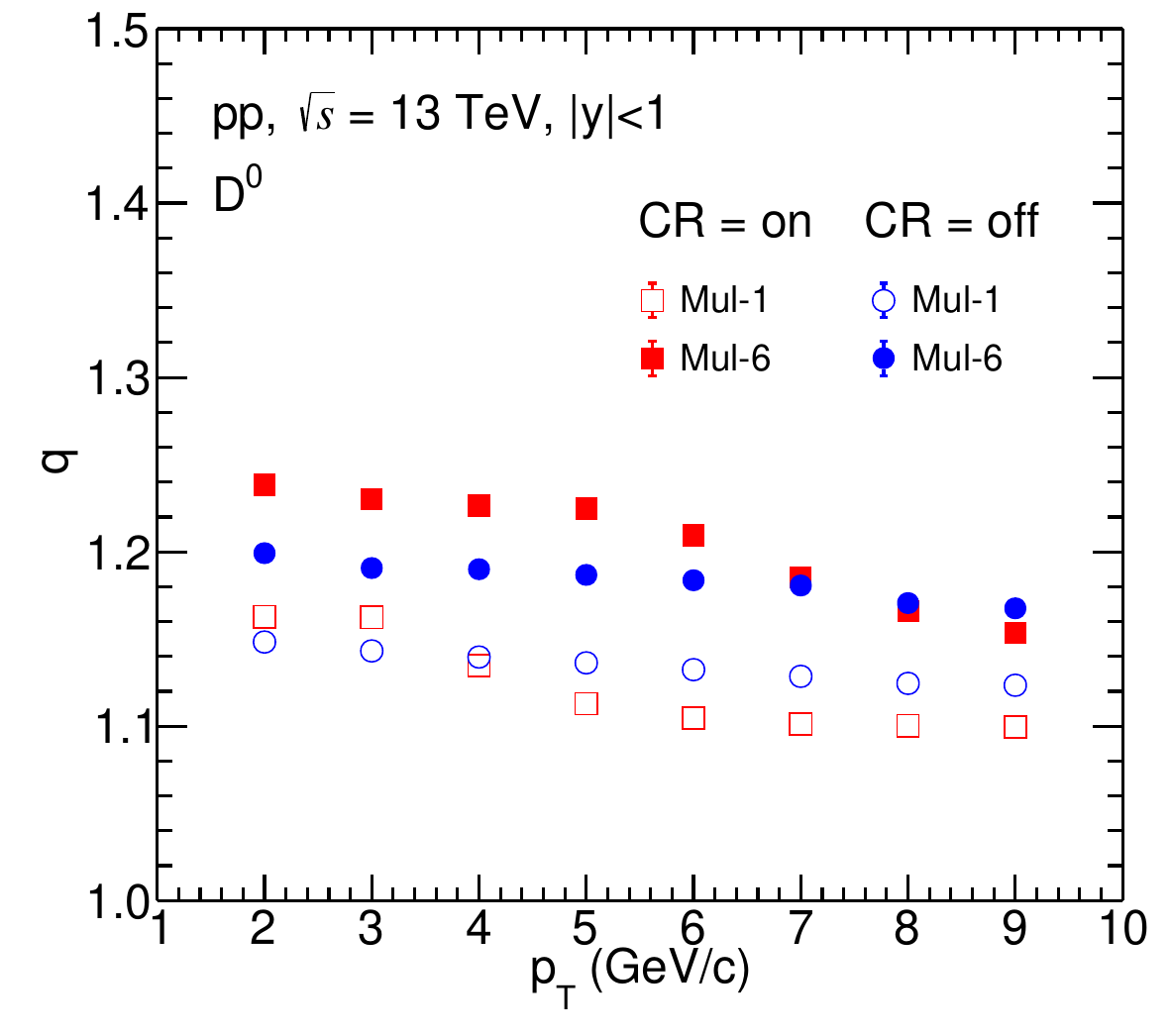} 
   
  \caption{(Color online) Upper and lower panels show the Tsallis extracted parameters, T and q with transverse momentum, respectively by considering both color reconnection modes on and off, for $\Lambda_{c}^{+} $ and $\rm{D^{0}}$. }
  \label{fig:3}
\ec
\end{figure*}

From the spectra, it is clearly observed that the particle production mechanism is different at different $p_{\rm{T}}$ regions (for instance, $p_{\rm{T}} <$ 3 GeV/c and above). From the ratio between MC simulated data points with the fit function, it is observed that non-extensive statistics provide a reasonably good description for open heavy flavor transverse momentum spectra for a broader $p_{\rm{T}}$ range. The fitting function fits relatively well with simulated data points within uncertainties for the  $p_{\rm{T}}$ $<$ 8.0 GeV/c except for $\Lambda_{c}^{+}$ in CR=off mode. In the context of multiplicity classes, we observed that the Tsallis function explains high-multiplicity classes relatively better than low-multiplicity classes at the high-$p_{\rm{T}}$ range.  We also observed the effect of color reconnection on open charm hadrons production dynamics using the tuning feature of PYTHIA8. The dependence of fitting parameters T and q  with charged-particle multiplicity, transverse momentum, and pseudorapidity are studied separately.
 
\begin{figure*}[!ht]

\bc
 
   \includegraphics[scale=0.33]{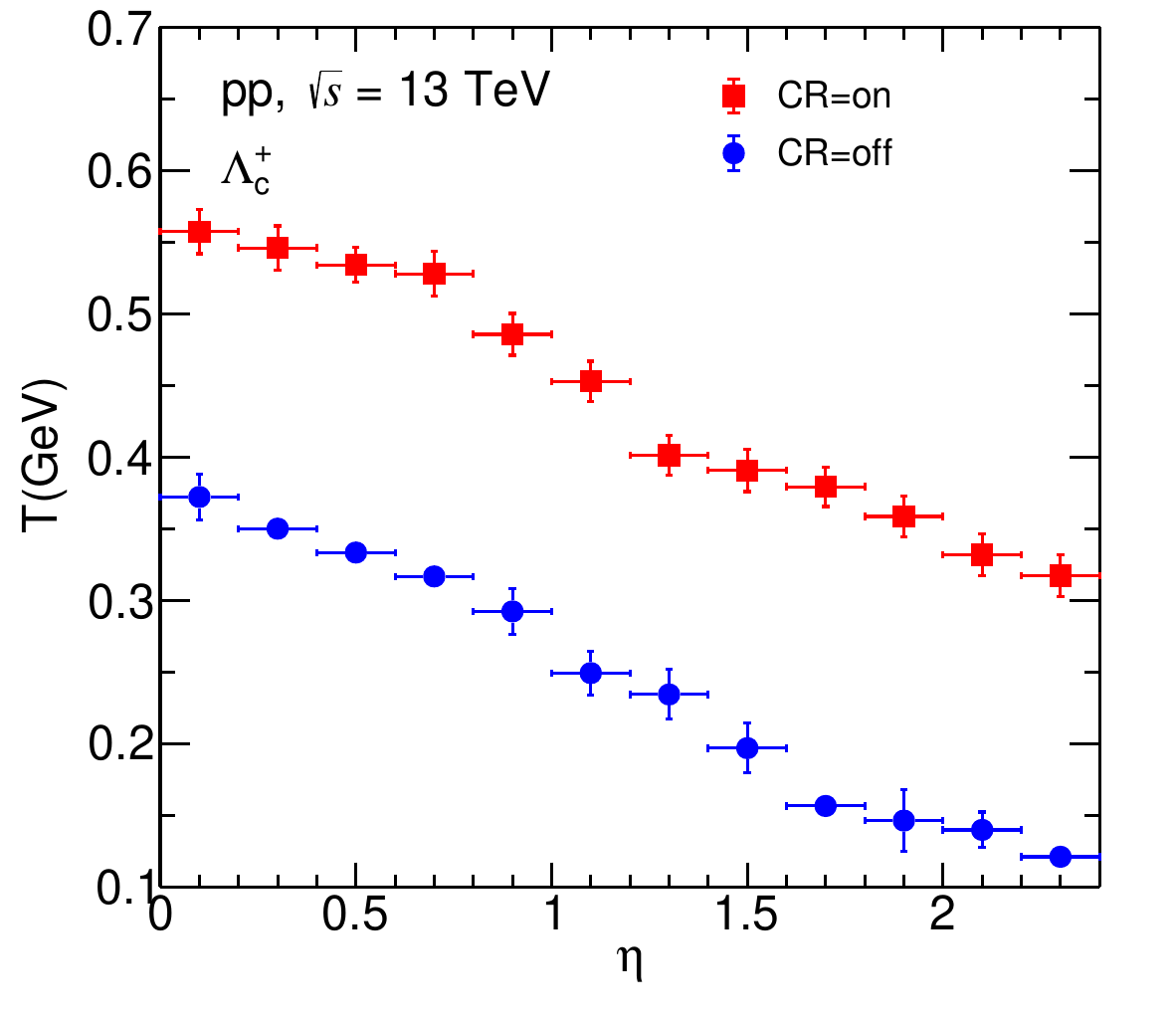}
   \includegraphics[scale=0.33]{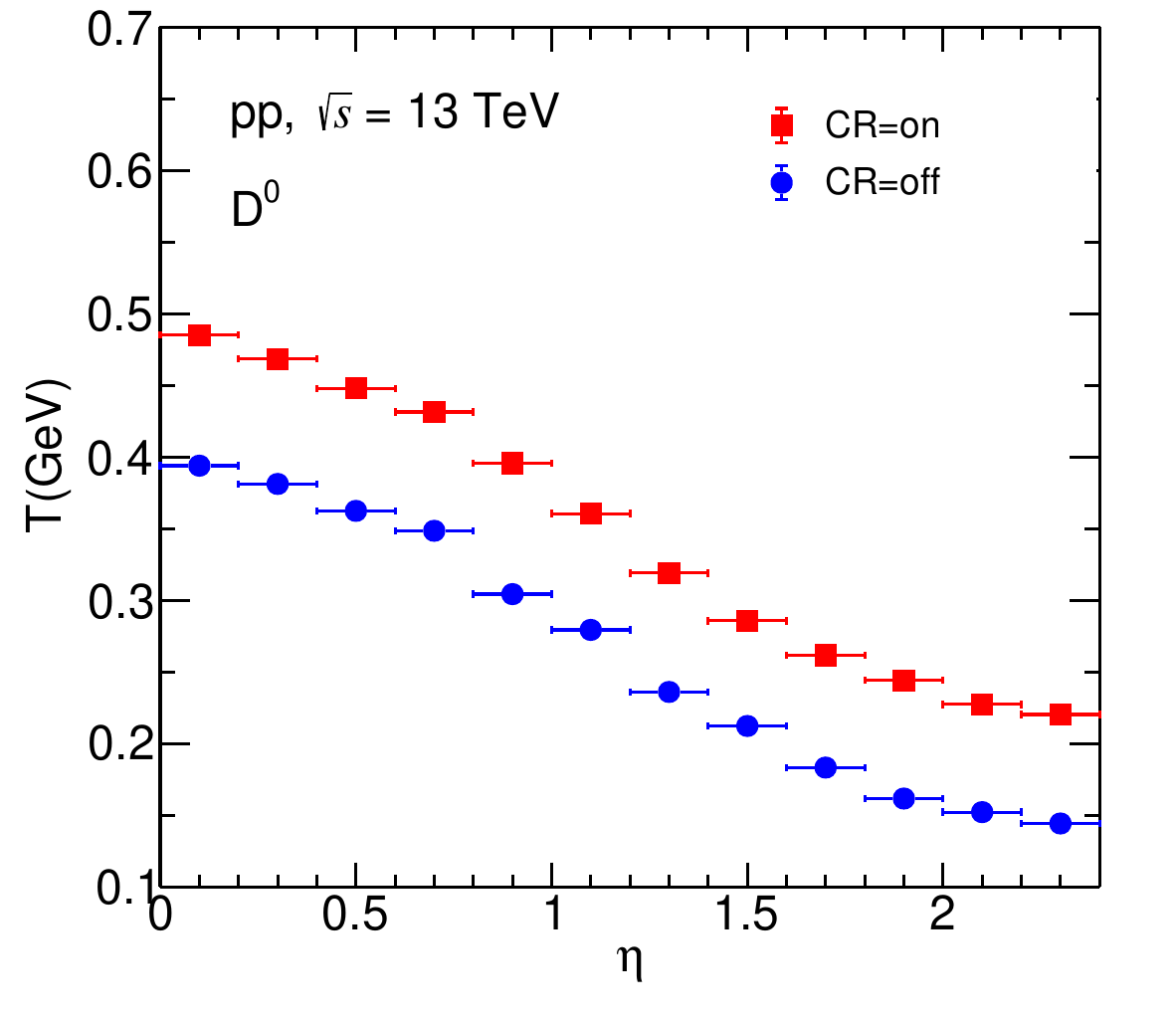}
   \includegraphics[scale=0.33]{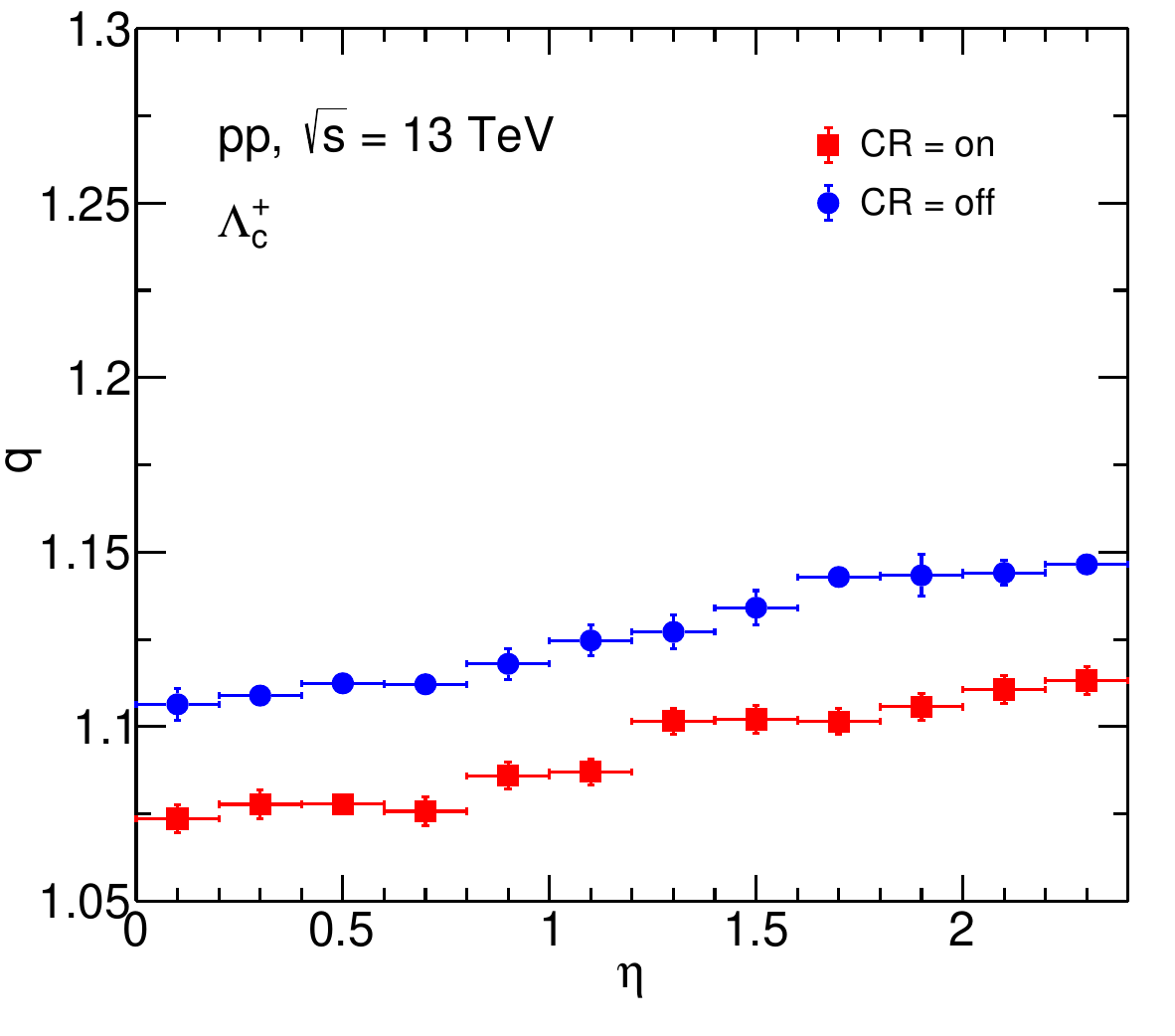}
   \includegraphics[scale=0.33]{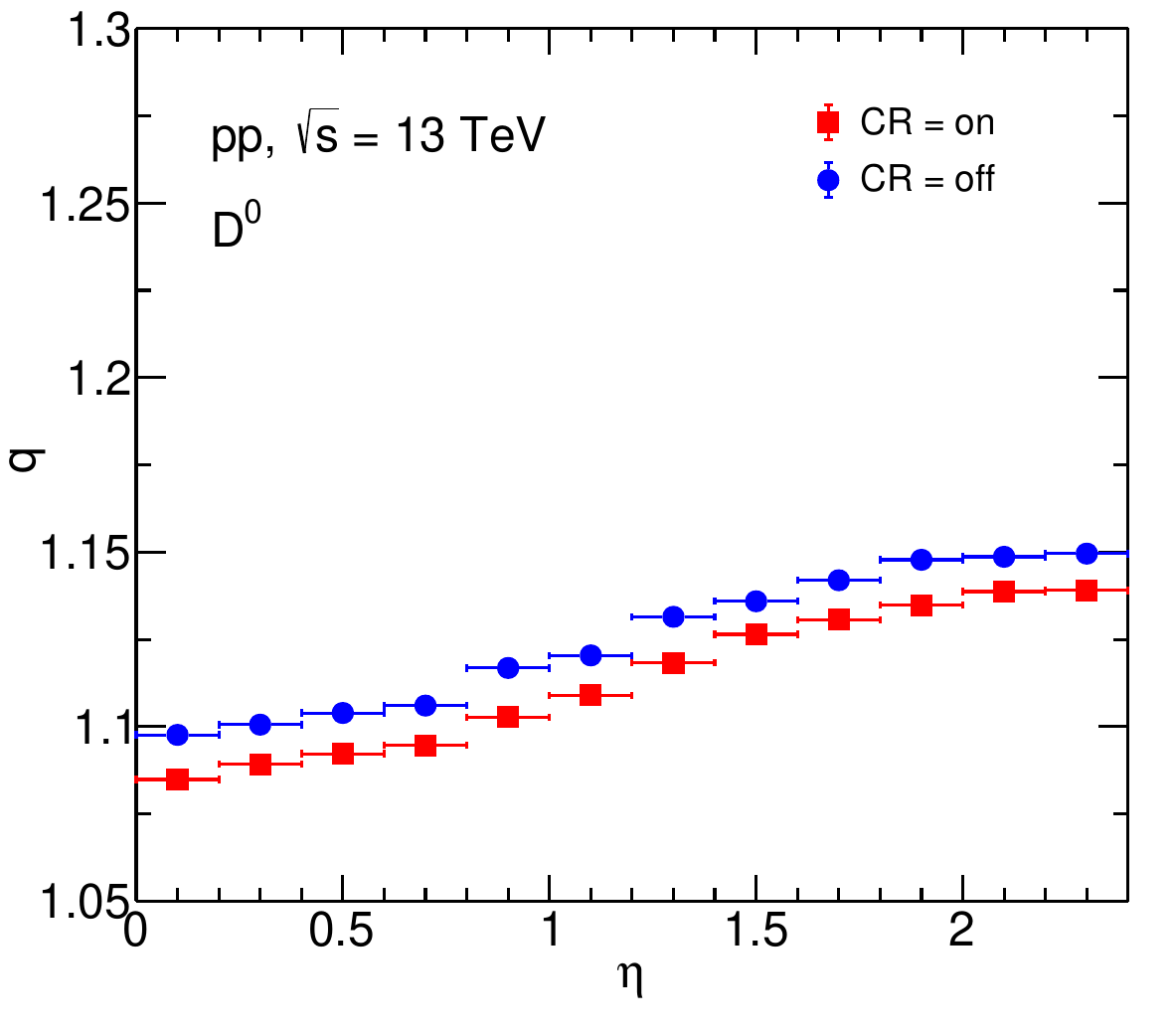}
  \caption{(Color online) Upper and lower panels show the Tsallis extracted parameters, T and q with pseudorapidity, respectively by considering both color reconnection mode on and off, for $\Lambda_{c}^{+} $ and $\rm{D^{0}}$. }
  \label{fig:4}
\ec
\end{figure*}

Figure~\ref{fig:2} shows that the extracted effective temperature (T) and the non-extensive parameter (q)  increase with charged-particle multiplicity class ($N_{ch}$) both for $\Lambda_{c}^{+}$ and $\rm{D^{0}}$ obtained from PYTHIA8. The obtained simulation results are compared with the ALICE experimental data by analyzing the multiplicity dependence of prompt $\Lambda_{c}^{+} $ and $\rm{D^{0}}$ production in pp collisions at $\sqrt{s}$ = 13 TeV with the Tsallis distribution  function~\cite{ALICE:2021npz}. Because of unavailability of the inclusive production of $\Lambda_{c}^{+} $ and $\rm{D^{0}}$  as a function of charged-particle multiplicity, here, we have compared with the multiplicity-dependent prompt $\Lambda_{c}^{+} $ and $\rm{D^{0}}$ results in pp collisions at $\sqrt{s}$ = 13 TeV. So, the ALICE experimental data comparison is only for illustration purposes. The extracted effective temperature from ALICE data is observed to increase with charged-particle multiplicity  ($N_{ch}$), while the non-extensive parameter is almost constant within the uncertainties both for $\Lambda_{c}^{+}$ and $\rm{D^{0}}$. 

This increase of T and q is observed for both the presence and absence of the color reconnection mode of hadronization implemented in PYTHIA8. Interestingly, it is observed that the effective temperature is less in color reconnection mode off as compared to the color reconnection mode on for all multiplicity classes. This could be due to the fact that CR along with MPIs could mimic the flow-like effects as discussed in Ref.~\cite{OrtizVelasquez:2013ofg}, and the absence of either of them in the simulation process is picked up by the decrease in effective temperature. CR being a hadronization
process, it increases the number of particles in the system, and along with MPI, it helps the system to approach a thermal-like description. However, it is observed that the effective temperature is lower for the CR off case, as compared to the CR on case, which goes in line with the observation that the $p_T$-spectra in CR off cases is harder than the CR on cases. When CR is off, the pQCD processes stay in the system making the system temperature lower than the case of CR on cases. The higher value of q with charged-particle multiplicity indicates that the particle is away from equilibrium in all multiplicity classes. Since $\Lambda_{c}^{+}$ and $\rm{D^{0}}$ are heavy particles, their production dynamics are different compared to light flavor particles studied in Ref.~\cite{Khuntia:2018znt}. The chances of equilibrium in the medium are very low for heavy-flavored particles. \\

Next, we attempted to study the behavior of T and q  with transverse momentum ($p_{\rm{T}}$), considering only the lowest and highest multiplicity classes. This is done by varying the upper limit of $p_{\rm{T}}$-spectra and keeping the lower limit fixed. We have chosen various $p_{\rm{T}}$ ranges starting from (0.2-2.0) GeV/c  to  (0.2-9.0) GeV/c with increasing a step of 1.0 GeV/c for each  $p_{\rm{T}}$ bin. In the $p_{\rm{T}}$-axis of Fig. \ref{fig:3}, the binning is 
taken at the upper limit of $p_{\rm{T}}$ range. The main idea of this study is to go on adding high-$p_{\rm{T}}$ particles to the system and study the change in the dynamics of the system. The Tsallis parameters T and q are extracted by fitting above mentioned $p_{\rm{T}}$ ranges with the Tsallis distribution function. A similar analysis is performed in Ref.~\cite{Patra:2020gzw} taking experimental data for charged pions and all charged-particles in heavy-ion collisions by varying the upper limit of $p_{\rm{T}}$ up to 50 GeV/c. The authors have studied the behavior of T and q with $p_{\rm{T}}$ for various collision energies and different centrality classes. 

In addition to that in Ref.~\cite{Rath:2019cpe}, the evolution of Tsallis parameter T and q with $p_{\rm{T}}$ is studied for pp collisions with experimental data for all charged-particles. In this work, we have studied the evolution of Tsallis parameters T and q with  $p_{\rm{T}}$ for $\Lambda_{c}^{+} $ and $\rm{D^{0}}$ up to $p_{\rm{T}}$ = 9.0 GeV/c in PYTHIA8. Since here we are using PYTHIA8 simulation data because of limitation in statistics at high-$p_{\rm{T}}$ we restrict the studies to $p_{\rm{T}}$ = 9.0 GeV/c. This study serves a dual purpose viz.,  because of the different number densities, the environment at low and high-multiplicity classes will be very different from each other. From the upper panel of  fig.~\ref{fig:3}, it is observed that the effective temperature as a function of  $p_{\rm{T}}$ shows an increasing trend with an increase in $p_{\rm{T}}$ both for the lowest and highest multiplicity class and it tends to saturate at high-$p_{\rm{T}}$ for  $\Lambda_{c}^{+}$.  The step like behaviour is found in T around $p_{\rm{T}} \simeq $ 6 GeV for $\Lambda_{c}^{+}$ with CR mode on could be due to a different particle production mechanisms being responsible for different $p_{\rm{T}}$ ranges. The contribution from hard pQCD processes and jets starts to dominate around $p_{\rm{T}}$ $\simeq$ 6 GeV/c, which may be responsible for the different behavior in Tsallis parameters as a function of $p_{\rm{T}}$. Interestingly, with the CR mode off case this saturation behavior of T as a function of $p_{\rm{T}}$ seems to disappear for $\Lambda_{c}^{+}$. However, for $\rm{D^{0}}$ which shows a slightly increasing trend at high $p_{\rm{T}}$ for both modes of color reconnection. 

 The non-extensive parameter q in the lower panel of fig.~\ref{fig:3}, it is observed that it decreases with $p_{\rm{T}}$ up to (6.0-7.0) GeV/c and then saturates towards high-$p_{\rm{T}}$ both for the lowest and highest multiplicity classes. Both $\rm{D^{0}}$ and $\Lambda_{c}^{+}$ were observed to follow the same trend irrespective of the mode of color reconnection. Similar to T, q also shows a change in the behavior as a function of $p_{\rm{T}}$ around $\simeq$ 6 GeV/c for $\Lambda_{c}^{+}$ and it follows the same explanation as above.   \\

We further extend our analysis to study the pseudorapidity ($\eta$) dependence of T and q  for the entire $p_{\rm{T}}$ range considering different $\eta$  bins from 0 to 2.4 in steps of 0.2 covering mid and forward rapidities. This is followed by extracting values of T and q for each $\eta$-bin after fitting the spectra with the Tsallis function. This is made possible because PYTHIA8 is quite robust in describing both the low and high-$p_{\rm{T}}$ part of the experimental transverse momentum spectra~\cite{Patra:2020gzw, Khuntia:2018znt,Rath:2019cpe}. \\

Fig.~\ref{fig:4} shows the pseudorapidity ($\eta$) dependence of effective temperature (T) and the non-extensive parameter (q). It is observed that the effective temperature decreases with an increase in pseudorapidity ($\eta$). A similar result was found in Ref~\cite{Ajaz:2022mga}. This observation of a higher effective temperature at midrapidity is because of the large transfer of energy density in the lower pseudorapidity region, which leads to a very high degree of excitation of the system. This results in a higher T-value in the lower pseudorapidity region. This also goes in line with a higher number of secondaries being produced at the midrapidity as compared to the forward rapidities. With a higher number of particles at the midrapidity region, it is expected that the system should approach a thermodynamic equilibrium faster than the system usually produced at the forward rapidity. This is also reflected in our results, where we observe the non-extensive parameter (q) being closer to unity for the midrapidity region and it goes further away from equilibrium when one moves towards the forward rapidities. This is in line with the results reported in Ref. ~\cite{Ajaz:2022mga}. \\
\begin{figure*}[!ht]

\bc
  \includegraphics[scale=0.29]{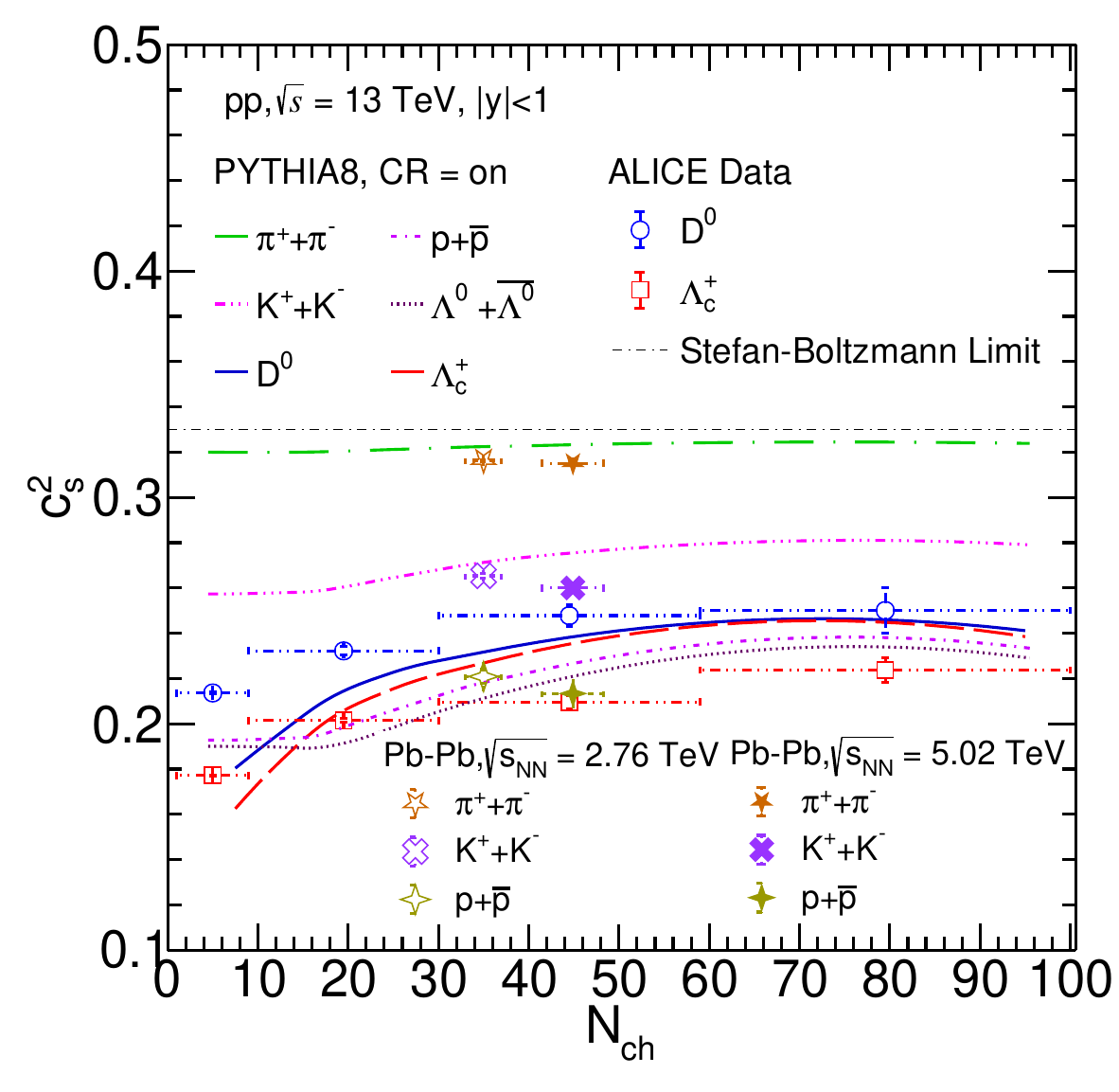}
  \includegraphics[scale=0.29]{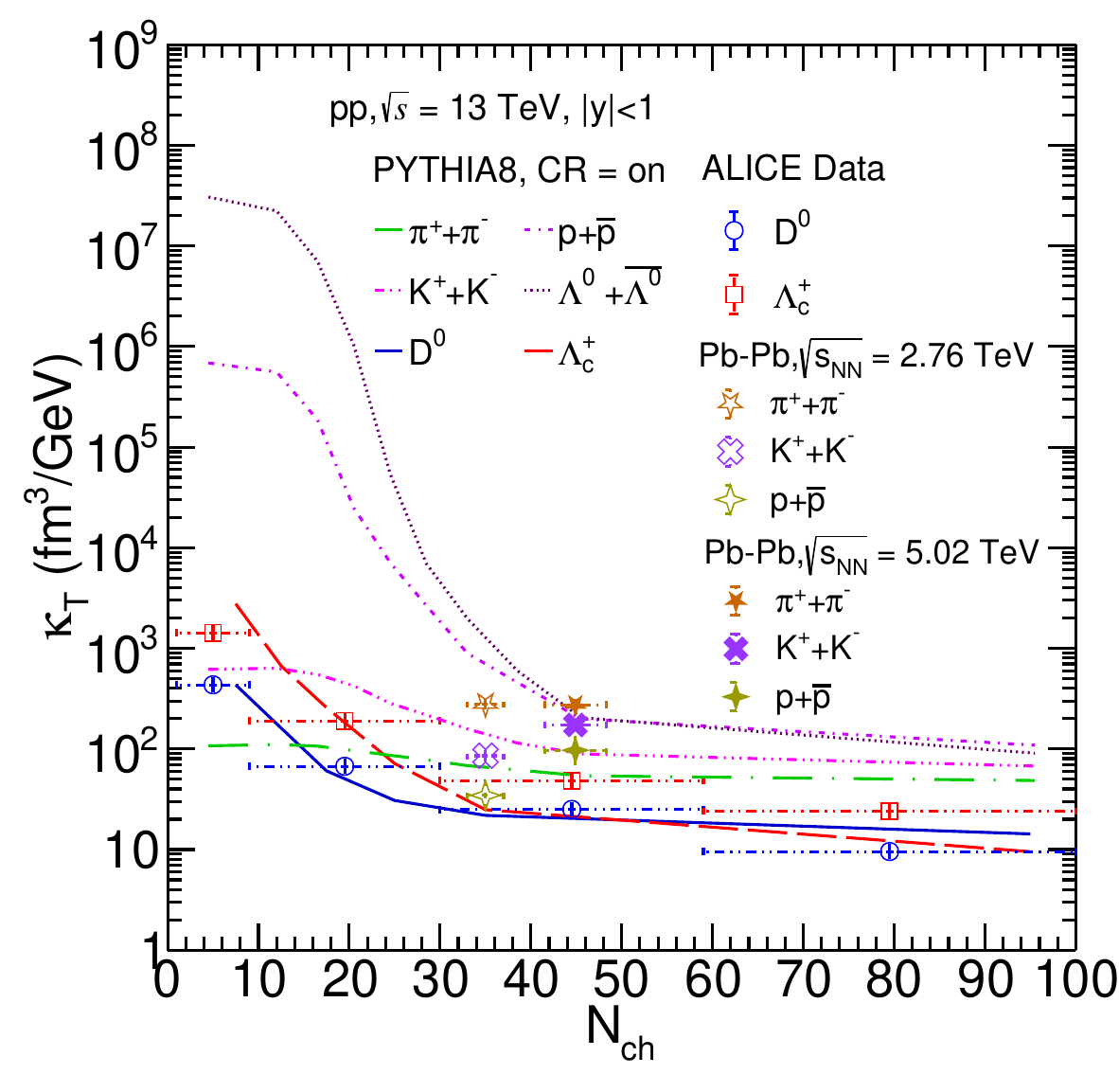} 
  \includegraphics[scale=0.29]{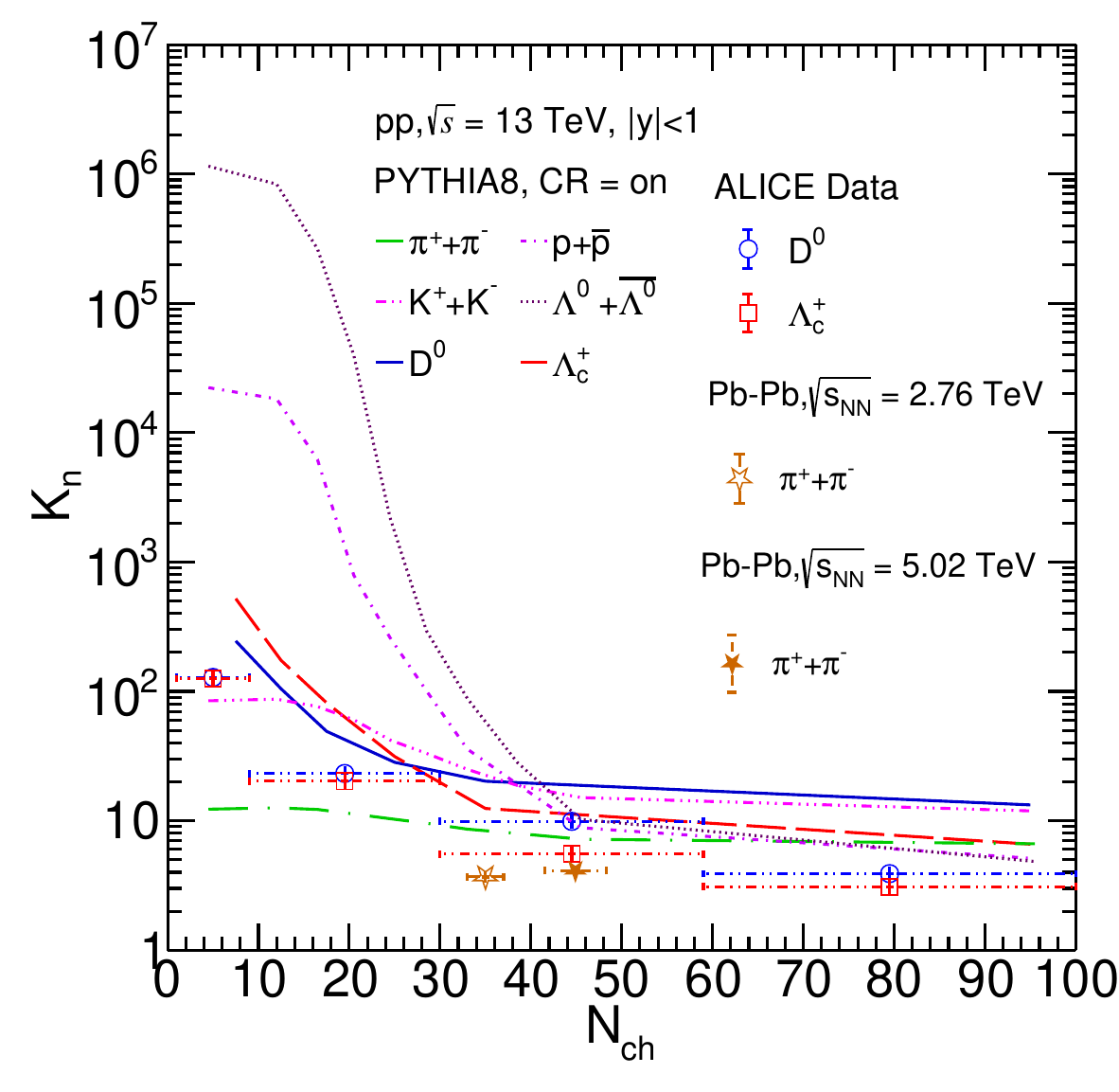}

  \caption{(Color online) Thermodynamic quantities like, the squared speed of sound  ($c_{s}^{2}$) (left), isothermal compressibility ($\kappa_{T}$) (middle), Knudsen number ($K_{n}$) (right), with charged-particle multiplicity for $\Lambda_{c}^{+} $ and $\rm{D^{0}}$ in pp collisions at $\sqrt{s}$ = 13 TeV using PYTHIA8. The obtained results are compared with the ALICE data for $\Lambda_{c}^{+} $ and $\rm{D^{0}}$ in pp collisions at $\sqrt{s}$ = 13 TeV ~\cite{ALICE:2021npz}. Further, the results are compared with the Pb-Pb collisions in (70-80) \% centrality class at $\sqrt{s_{NN}}$ = 2.76 TeV ~\cite{ALICE:2015dtd}and $\sqrt{s_{NN}}$ = 5.02 TeV ~\cite{ALICE:2019hno} for identified particles, taking the Tsallis parameters reported in Ref.~\cite{Patra:2020gzw}. The results are compared with light and strange particle results obtained in pp collisions at  $\sqrt{s}$ = 13 TeV using PYTHIA8 and reported in Ref.~\cite{Deb:2020}.}
  \label{fig:5}
\ec
\end{figure*}

\subsection{Multiplicity dependence of $c_{s}^{2}$, $\kappa_{T}$ and $K_{n}$}
\label{subsection_2}

Here, we discuss the variation of thermodynamic quantities like the squared speed of sound ($c_{s}^{2}$), isothermal compressibility ($\kappa_{T}$), the Knudsen number($K_{n}$)  with charged-particle multiplicity, which is briefly motivated in section~\ref{formalism}. Using the Tsallis distribution function given by Eq.~\ref{eq_2}, these thermodynamical quantities are calculated and the final mathematical forms are expressed in Eqs.~\ref{eq_11}~\ref{eq_12} and ~\ref{eq_13}. The Tsallis parameters, namely, T and q obtained as a function of charged-particle multiplicity in the previous section~\ref{pt_spectra} are used to compute the above-mentioned thermodynamic quantities for $\Lambda_{c}^{+}$ and $\rm{D^{0}}$. It is worth noting that the (T, q) used here are from CR = on mode only as CR with MPI was found to mimic flow-like features in a small system, as reported in Ref.~\cite{OrtizVelasquez:2013ofg}. \\

Figure~\ref{fig:5} shows the variation of $c_{s}^{2}$, $\kappa_{T}$ and $K_{n}$  with charged-particle multiplicity for $\Lambda_{c}^{+}$ and $\rm{D^{0}}$. We observed that $c_{s}^{2}$ increases with charged-particle multiplicity both for $\Lambda_{c}^{+}$ and $\rm{D^{0}}$, although the $c_{s}^{2}$ value remain away from the ideal gas limit or Stefan-Boltzmann limit (1/3), where the particles are assumed to be massless and non-interacting. But, this observation indicates that the high-multiplicity medium is more thermal-like than the low-multiplicity environment. Since in the present work, we are dealing with heavy flavor particles, which have comparatively higher mass compared to light flavors, it is an interesting observation that the effect of interaction strength is also prevalent in the charm sector. We further observed that the mass ordering in the speed of sound is also preserved in all multiplicity classes, i.e., massive particles ($\Lambda_{c}^{+}$) have less $c_s^{2}$ value as compared to $\rm{D^{0}}$, which is relatively less massive. This result agrees with previous findings reported in Ref.~\cite{Deb:2019yjo}. In the case of $\kappa_{T}$, we observe that it is very sensitive to charged-particle multiplicities in the final state. In lower multiplicity classes, the value of isothermal compressibility is relatively high. It starts to decrease with the increase in particle multiplicity until it attains a minimum value at the higher multiplicity classes. This observation shows that the system is more incompressible at higher multiplicities. This result also agrees with previous findings, as shown in Ref.~\cite{Sahu:2020swd}. Further, the Knudsen number is observed to decrease with charged-particle multiplicity. Although the value of the Knudsen number remains above unity for all multiplicity classes and hence the applicability of hydrodynamic treatment is not possible. However, owing to comparatively less value of $K_{n} $ towards high-multiplicity indicates the probability of high-multiplicity region being relatively thermalized than low-multiplicity.

Additionally, the thermodynamic quantities such as $c_{s}^{2}$, $\kappa_{T}$, and $K_{n}$ obtained from PYTHIA8 simulation are compared with the analyzed ALICE data for prompt $\Lambda_{c}^{+}$ and $\rm{D^{0}}$ in pp collisions at $\sqrt{s}$ = 13 TeV using the extracted Tsallis parameters as shown in Fig.~\ref{fig:2}. Furthermore, $c_{s}^{2}$, $\kappa_{T}$, and $K_{n}$ obtained for open heavy flavor hadrons are compared with the light and strange particles results in pp collisions at $\sqrt{s}$ = 13 TeV studied using PYTHIA8 and reported in one of our previous paper~\cite{Deb:2020}. The obtained thermodynamic results are compared with the Pb-Pb collisions in (70-80) \% centrality class (designated by $N_{\rm ch} \equiv <dN_{\rm ch}/d\eta>$) at $\sqrt{s_{NN}}$ = 2.76 TeV ~\cite{ALICE:2015dtd} and $\sqrt{s_{NN}}$ = 5.02 TeV ~\cite{ALICE:2019hno} for identified particles, taking the Tsallis parameters reported in Ref.~\cite{Patra:2020gzw}. \\ 

It is evident to mention that the values of T and q obtained from section~\ref{pt_spectra} are different for  $\Lambda_{c}^{+}$ and $\rm{D^{0}}$. Hence, we consider a differential freeze-out scenario, in which higher mass particles decouple from the system early in time indicating a higher Tsallis effective temperature parameter. These particles are expected to carry more initial non-equilibrium effects. Therefore the thermodynamics quantities such as $c_{s}^{2}$, $\kappa_{T}$, and $K_{n}$ are particle species dependent. Again, from the above three observations, it is clear that the system shows less thermodynamic behavior at low multiplicities as compared to higher multiplicity classes. 

\subsection{Correlation between various quantities}
\label{subsection_3}

The medium formed in the hadronic or nucleus-nucleus collisions at the ultra-relativistic energies is very complex in nature. It involves the interplay of various processes like underlying events, MPIs, rescattering, etc. Some processes are significant in the partonic phase, while others dominate in the hadronic phase. It is possible to expect some reminiscence of partonic level effects in the hadron gas. With this motivation, in this section, we discuss the correlation between Tsallis effective temperature and non-extensive parameter as a function of charged-particle multiplicity and inverse of Knudsen number with the number of multi-partonic interactions ($\rm{n_{MPI}}$). For simplicity, we call the inverse of the Knudsen number as inverse ${K_{n}}$ from now onwards.

\begin{figure*}[!ht]

\bc
  \includegraphics[scale=0.33]{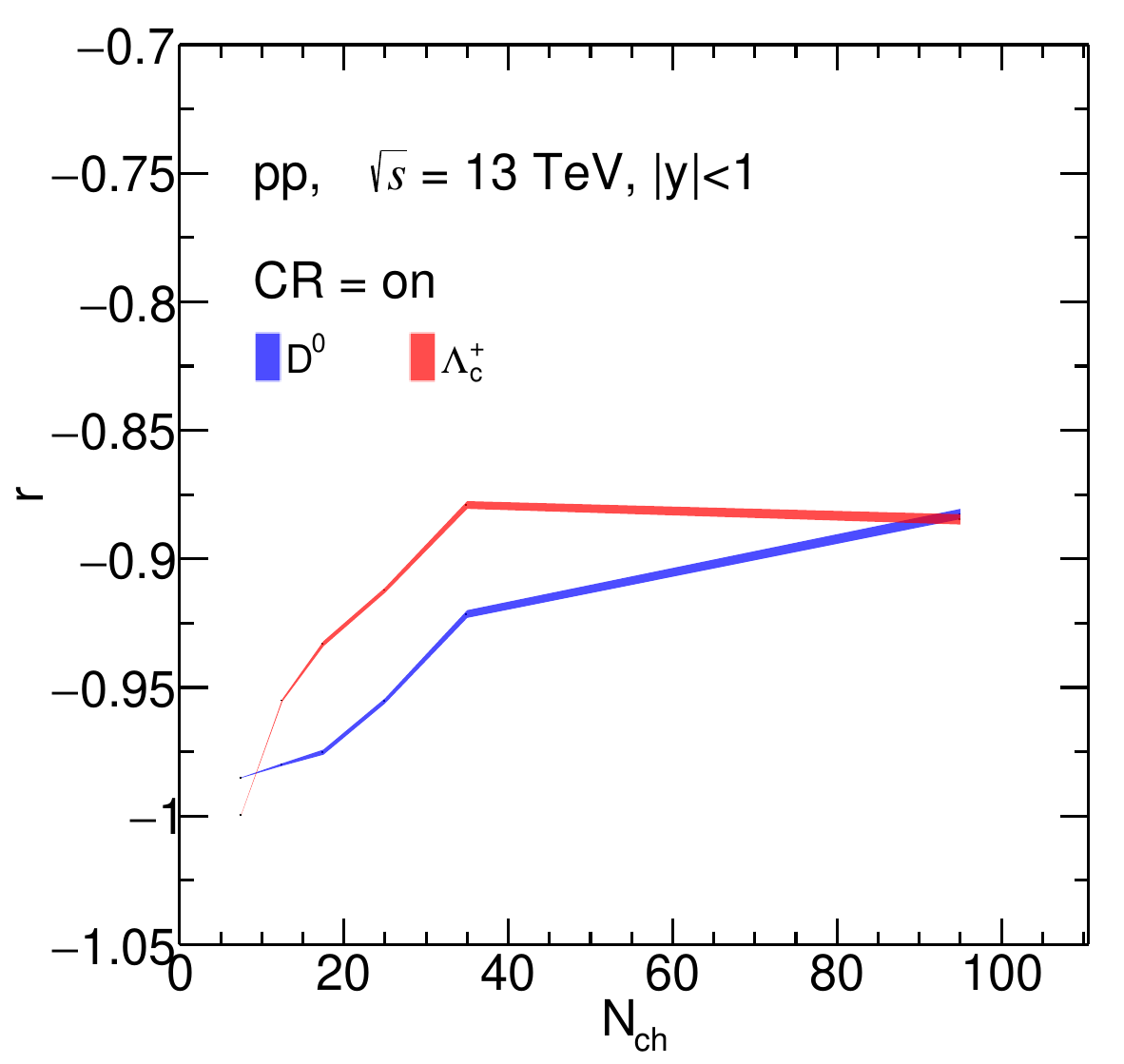}
  \includegraphics[scale=0.33]{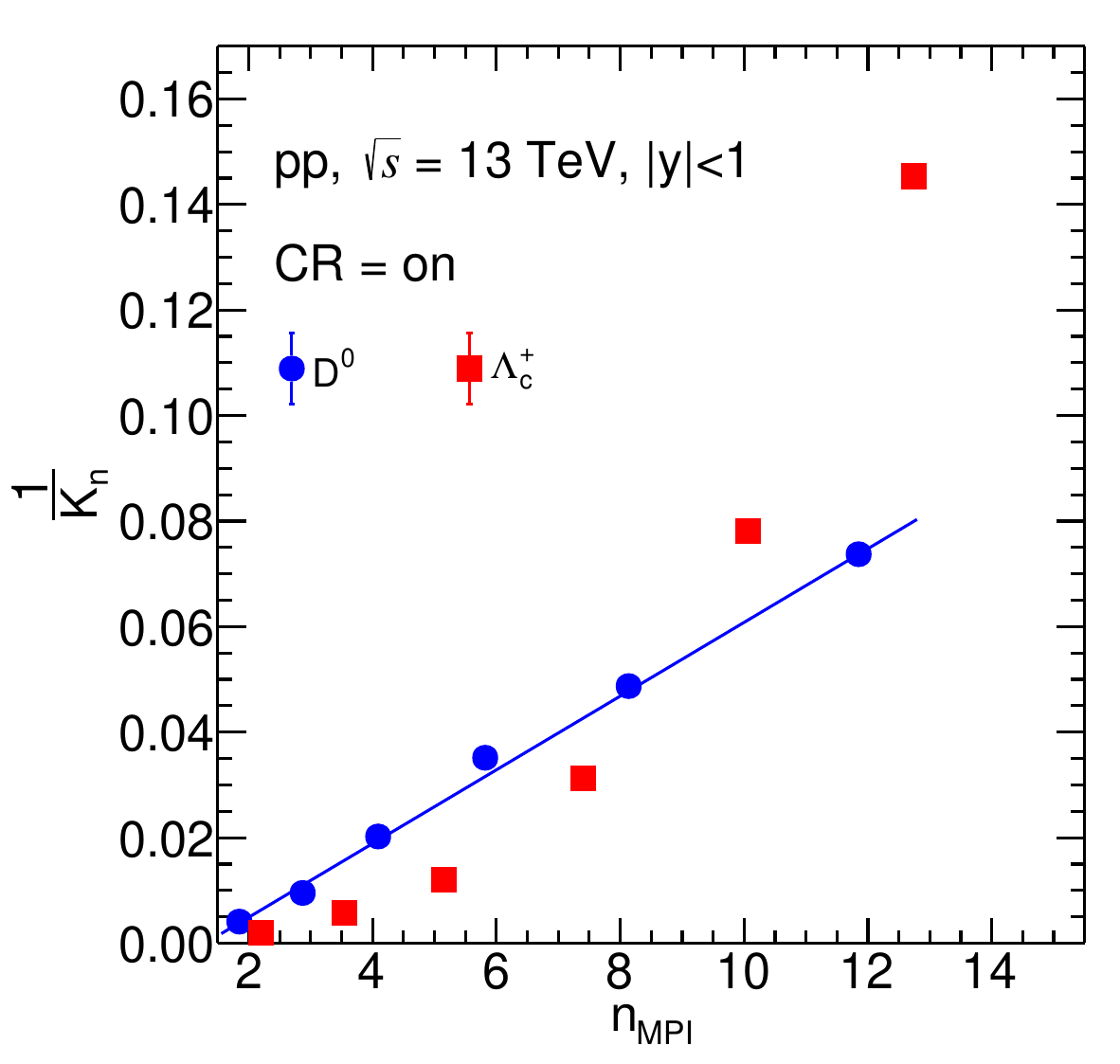} 
  
   \caption{(Color online) Correlation between (left) Tsallis parameters as a function of charged-particle multiplicity and (right) inverse Knudsen number and $\rm{n_{MPI}}$.}

  \label{fig:6}
\ec
\end{figure*}

\subsubsection{Correlation between Tsallis parameters}

In section~\ref{pt_spectra}, we observed that the Tsallis effective temperature (T) is related to the non-extensive parameter (q). Further, it is observed that the q values deviate from unity in the case of a system following the Tsallis-Boltzmann distribution. This may indicate the presence of long-range correlations, temperature fluctuation, and finiteness of the system. However, in Ref.~\cite{Deb:2019yjo}, it is observed that the finite size effect alone cannot explain the appearance of a deviation from q = 1, which suggests the presence of correlations in the QCD medium plays a vital role in the onset of non-extensivity. Here, we explicitly observe the variation in T with the q-parameter as a function of charged-particle multiplicity. We have extracted various values of T and q from the $p_{\rm{T}}$-spectra by choosing 10 different $p_{\rm{T}}$ ranges: (0.2-9.0), (0.4-8.5), (0.6-8.2), (0.8-7.5), (1.0-7.0), (1.5-6.5), (2.0-6.0), (1.5-8.5), (2.2-8.5)  and (3.0-9.0) GeV/c  for each multiplicity. We have chosen $p_{\rm{T}}$ ranges in such a way that both thermal and/or power-law contributions are present. We have obtained the correlation between T and q parameters by using the following relation:\\
 
 \begin{equation}
r(T,q) =\frac{Cov(T,q)}{\sigma_{T}\sigma_{q}}, 
\label{eq_15}
\end{equation}
 
where $\sigma_{T}$ is the standard deviation corresponding to T and $\sigma_{q}$ is the standard deviation corresponding to q. Cov(T,q) is the covariance of T and q. r(T,q) denotes the correlation coefficient. 

For a given multiplicity, the standard deviation of T and q parameters is obtained using these 10 different values of T and q for the abovementioned $p_{\rm{T}}$-interval. The covariance between T and q is obtained for the same dataset. Finally, the correlation between T and q is obtained using Eq.~\ref{eq_15}. The same procedure is applied to each multiplicity class both for $\Lambda_{c}^{+}$ and $D^{0}$.

The left side of Fig.~\ref{fig:6} shows the correlation between T and q as a function of charged-particle multiplicity. It is observed that the correlation between T and q is negative, which means T and q are anti-correlated with each other. An increase in the T value decreases the q value and vice-versa. This anti-correlation is maintained for all the multiplicity classes. At the lowest multiplicity, T and q are perfectly anti-correlated, and this correlation decreases towards higher multiplicities. The uncertainty band in the figure
takes care of the uncertainties in the associated variables used to estimate the correlation coefficient.

\subsubsection{Correlation between $\frac{1}{K_{n}}$ and  $\rm{n_{MPI}}$ }

$\rm{n_{MPI}}$ gives information about the number of multi-partonic interactions for a given impact parameter in a given collision. If the number of multi-partonic interactions is relatively more, the system tends toward thermal equilibrium. Similar observation can be drawn from the inverse ${K_{n}}$. A large value of the inverse ${K_{n}}$ indicates that the system is closer to a thermal-like system, as evident from Fig.~\ref{fig:5}. Here, we tried to explore the existence of any correlation between these two quantities and how this correlation behaves for a heavy-flavor baryon and meson. From section~\ref{subsection_2}, we found the information about the variation of the inverse ${K_{n}}$ with charged-particle multiplicity. MC model PYTHIA8 provides information about the $\rm{n_{MPI}}$ in each multiplicity class. With the knowledge of both these quantities with charged-particle multiplicity, we have obtained the correlation between the inverse ${K_{n}}$ and $\rm{n_{MPI}}$ as shown on the right side of Fig.~\ref{fig:6}. It is observed that there is a perfect correlation (or linear dependence) between the inverse ${K_{n}}$ with $\rm{n_{MPI}}$ for $\rm{D^{0}}$. However, it is observed that the dependence of inverse ${K_{n}}$ on $\rm{n_{MPI}}$ departs from linear dependence for $\Lambda_{c}^{+}$. This implies that the number of parton-parton interactions varies differently for a baryon as compared to a meson because of their different quark constituents. That means the meson carries the initial state information more accurately as compared to the baryon, because of a 
nice linear behavior with $\rm{n_{MPI}}$. It is an interesting observation that needs to be explored further to draw a better conclusion on the production dynamics of heavy-flavored baryons and mesons, and baryons and mesons, in general.

\section{SUMMARY}
\label{summary}

In this work, we have studied the heavy-flavored hadrons, namely $\Lambda_{c}^{+}$ and $\rm{D^{0}}$ production dynamics using a pQCD-inspired model called PYTHIA8 in pp collisions at $\sqrt{s}$ = 13 TeV. 
The important observations of this paper are summarized below.
\begin{enumerate}

 \item The transverse momentum ($ p_{\rm T}$) spectra of $\Lambda_{c}^{+}$ and $\rm{D^{0}}$ produced in pp collision at the center of mass energy $\sqrt{s}$ = 13 TeV are studied using simulated data. These $ p_{\rm T}$ spectra provide useful information regarding the production mechanism as well as freeze-out conditions of $\Lambda_{c}^{+}$ and $\rm{D^{0}}$. We have analyzed the $ p_{\rm T}$ spectra of $\Lambda_{c}^{+}$ and $\rm{D^{0}}$  with the thermodynamically consistent Tsallis distribution function up to  $p_{\rm{T}}$ = 9 GeV/c. The effect of color reconnection on particle production is clearly observed.

 \item The variation of effective temperature (T) and non-extensive parameter (q) for $\Lambda_{c}^{+}$ and $\rm{D^{0}}$, extracted from Tsallis distribution function is studied with charged particle multiplicity ($N_{ch}$),  transverse momentum spectra ($p_{\rm{T}}$) and pseudorapidity ($\eta$). The $N_{ch}$ dependence study of T and q provides insight into the dynamics of the system produced with different densities in pp collisions. Meanwhile, the study of $p_{\rm{T}}$ dependence of T and q sheds light on particle production mechanisms, and the rapidity dependence covers the phase-space analysis of the particles produced in hadronic collisions.
 
 \item It is found that for heavy-flavored hadrons, both the effective temperature and non-extensive parameter increase with charged-particle multiplicity. However, the effective temperature increases with $p_{\rm{T}}$ both for low and high-multiplicity classes. On the other hand, the non-extensive parameter, q decreases with an increase in $p_{\rm{T}}$ and then shows a saturation behavior towards high-$p_{\rm{T}}$ both for low and high-multiplicity classes. However, for the high-multiplicity classes, there is a considerable change in saturation behavior. The effective temperature decreases and the non-extensive parameter, q increases while going from midrapidity to the forward rapidity region both for $\Lambda_{c}^{+}$ and $\rm{D^{0}}$.

 \item  To explore the possible thermal behavior of the produced system, the squared speed of sound, isothermal compressibility, and the Knudsen number are studied as a function of charged-particle multiplicity. It is observed that at high-multiplicity, the system behaves relatively more like a thermalized medium. 

 \item We have observed an anti-correlation relation between Tsallis Parameters viz., T and q with charged-particle multiplicity. Additionally, we observe a strong correlation between the inverse of the Knudsen number and the number of multi-partonic interactions for meson. The correlation seems to break for heavy-flavored baryon. This needs further investigation to see any possible baryon-meson effect at the partonic level activities in the process of particle production. \\

 \item It would be intriguing to investigate the multiplicity dependence of heavy flavor production in pp collisions with precise vertex measuring capability and the high statistics data of Run 3 and Run 4 setup of ALICE 2. This may reveal many exciting physics results which help to constrain the theoretical models to a good extent.
 
 \end{enumerate}

\section*{ACKNOWLEDGEMENT}
BS acknowledges the doctoral fellowship from CSIR, Government of India. The authors gratefully acknowledge the DAE-DST, Govt. of India funding under the mega-science project -- “Indian participation in the ALICE experiment at CERN" bearing Project No. SR/MF/PS-02/2021-IITI (E-37123). The usage of the ALICE Tier-3 computing facility at IIT Indore is gratefully acknowledged.


\section*{Data Availability Statement}
This paper uses PYTHIA simulation data, which could be made available with a reasonable request to the corresponding author.

\end{document}